\documentclass[twocolumn, twocolappendix]{aastex63}

\usepackage{xcolor}

\def\tractor{\texttt{The~Tractor}}
\def\farmer{\texttt{The~Farmer}}
\def\Euclid{\textit{Euclid}}

\usepackage{amsmath}

\newcommand{\likelihood}{\mathcal{L}}
\newcommand{\dd}{\partial}
\newcommand{\var}{\textrm{var}}

\received{\today}
\revised{\today}
\accepted{\today}
\submitjournal{ApJS}

\shorttitle{The Farmer}
\shortauthors{Weaver et al.}
\graphicspath{{./figures}}

\begin{document}

\title{\texttt{The Farmer}:\\A reproducible profile-fitting photometry package for deep galaxy surveys}

\correspondingauthor{John R. Weaver}
\email{john.weaver.astro@gmail.com}

\author[0000-0003-1614-196X]{J. R. Weaver}
\affil{Cosmic Dawn Center (DAWN)}
\affil{Niels Bohr Institute, University of Copenhagen, Jagtvej 128, 2200 Copenhagen, Denmark}
\affil{Department of Astronomy, University of Massachusetts, Amherst, MA 01003, USA}

\author[0000-0001-5680-2326]{L. Zalesky}
\affil{Institute for Astronomy, University of Hawaii, 2680 Woodlawn Drive, Honolulu, HI 96822, USA}
\affil{Niels Bohr Institute, University of Copenhagen, Jagtvej 128, 2200 Copenhagen, Denmark}

\author[0000-0002-5588-9156]{V. Kokorev}
\affil{Kapteyn Astronomical Institute, University of Groningen, PO Box 800, 9700 AV Groningen, The Netherlands}
\affil{Cosmic Dawn Center (DAWN)}
\affil{Niels Bohr Institute, University of Copenhagen, Jagtvej 128, 2200 Copenhagen, Denmark}
 
\author[0000-0003-0639-025X]{C. J. R. McPartland}
\affil{Cosmic Dawn Center (DAWN)}
\affil{Institute for Astronomy, University of Hawaii, 2680 Woodlawn Drive, Honolulu, HI 96822, USA}
\affil{Niels Bohr Institute, University of Copenhagen, Jagtvej 128, 2200 Copenhagen, Denmark}
 
\author[0000-0003-3691-937X]{N. Chartab}
\affiliation{The Observatories of the Carnegie Institution for Science, 813 Santa Barbara St., Pasadena, CA 91101, USA}
\affiliation{Department of Physics and Astronomy, University of California, Irvine, CA 92697, USA}

\author[0000-0003-4196-5960]{K. M. L. Gould}
\affil{Cosmic Dawn Center (DAWN)}
\affil{Niels Bohr Institute, University of Copenhagen, Jagtvej 128, 2200 Copenhagen, Denmark}

\author[0000-0002-7087-0701]{M. Shuntov}
\affil{Institut d'Astrophysique de Paris, UMR 7095, CNRS,
    and Sorbonne Universit\'e, 98 bis boulevard Arago, 75014 Paris, France}

\author[0000-0002-2951-7519]{I. Davidzon}
\affil{Cosmic Dawn Center (DAWN)}
\affil{Niels Bohr Institute, University of Copenhagen, Jagtvej 128, 2200 Copenhagen, Denmark}

\author[0000-0002-9382-9832]{A. Faisst}
\affil{Infrared Processing and Analysis Center, California Institute of Technology, 1200 E. California Blvd, Pasadena, CA, 91125, USA}

\author{N. Stickley}
\affil{Infrared Processing and Analysis Center, California Institute of Technology, 1200 E. California Blvd, Pasadena, CA, 91125, USA}

\author{P.~L. Capak}
\affil{Cosmic Dawn Center (DAWN)}

\author[0000-0003-3631-7176]{S. Toft}
\affil{Cosmic Dawn Center (DAWN)}
\affil{Niels Bohr Institute, University of Copenhagen, Jagtvej 128, 2200 Copenhagen, Denmark}

\author{D. Masters}
\altaffiliation{NASA Postdoctoral Program (NPP) Fellow}
\affil{Jet Propulsion Laboratory, California Institute of Technology 4800 Oak Grove Drive Pasadena, CA 91109, USA}
\affil{Infrared Processing and Analysis Center, California Institute of Technology, 1200 E. California Blvd, Pasadena, CA, 91125, USA}

\author[0000-0001-5846-4404]{B. Mobasher}
\affil{Department of Physics and Astronomy, University of California, Riverside, 900 University Avenue, Riverside, CA 92521, USA}

\author{D. B. Sanders}
\affil{Institute for Astronomy, University of Hawaii, 2680 Woodlawn Drive, Honolulu, HI 96822, USA}

\author{O. B. Kauffmann}
\affil{Aix Marseille Univ, CNRS, LAM, Laboratoire d'Astrophysique de Marseille, Marseille, France}

\author[0000-0002-9489-7765]{H. J. McCracken}
\affil{Institut d'Astrophysique de Paris, UMR 7095, CNRS,
    and Sorbonne Universit\'e, 98 bis boulevard Arago, 75014 Paris, France}

\author[0000-0002-7303-4397]{O. Ilbert}
\affil{Aix Marseille Univ, CNRS, LAM, Laboratoire d'Astrophysique de Marseille, Marseille, France}

\author[0000-0003-2680-005X]{G. Brammer}
\affil{Cosmic Dawn Center (DAWN)}
\affil{Niels Bohr Institute, University of Copenhagen, Jagtvej 128, 2200 Copenhagen, Denmark}

\author{A. Moneti}
\affil{Institut d'Astrophysique de Paris, UMR 7095, CNRS,
    and Sorbonne Universit\'e, 98 bis boulevard Arago, 75014 Paris, France}




\begin{abstract}

While space-borne optical and near-infrared facilities have succeeded in delivering a precise and spatially resolved picture of our Universe, their small survey area is known to under-represent the true diversity of galaxy populations. Ground-based surveys have reached comparable depths but at lower spatial resolution, resulting in source confusion that hampers accurate photometry extractions. What once was limited to the infrared regime has now begun to challenge ground-based ultra-deep surveys, affecting detection and photometry alike. Failing to address these challenges will mean forfeiting a representative view into the distant Universe. We introduce \farmer{}: an automated, reproducible profile-fitting photometry package that pairs a library of smooth parametric models from \tractor{} (Lang et al. 2016) with a decision tree that determines the best-fit model in concert with neighboring sources. Photometry is measured by fitting the models on other bands leaving brightness free to vary. The resulting photometric measurements are naturally total, and no aperture corrections are required. Supporting diagnostics (e.g. $\chi^2$) enable measurement validation. As fitting models is relatively time intensive, The Farmer is built with high-performance computing routines. We benchmark \farmer{} on a set of realistic COSMOS-like images and find accurate photometry, number counts, and galaxy shapes. \farmer{} is already being utilized to produce catalogs for several large-area deep extragalactic surveys where it has been shown to tackle some of the most challenging optical and near-infrared data available, with the promise of extending to other ultra-deep surveys expected in the near future. \farmer{} is available to download from \href{https://github.com/astroweaver/the_farmer}{GitHub} and \href{https://doi.org/10.5281/zenodo.8205817}{Zenodo}.

\end{abstract}



\section{Introduction}
For most of its history, astronomy has been defined by the use of electro-magnetic waves to measure sources detected in the night sky. What began as a purely visual study was transformed in the late 19\textsuperscript{th} century with the advent of photographic plates that enabled precise observations from which the brightness of sources could be measured \citep{1888BuAsI...5..303B}. It was with such comparatively primitive technology that the first variable stars in Andromeda were identified, leading to the discovery of the `Island Universes' and later the expansion of the Universe \citep{Hubble1926, Hubble1929}. Now almost a century later, all scientific astronomical observations are captured on Charge-Coupled Devices, or CCDs \citep{Lesser_CCD}, further enhancing the accuracy and precision of photometry.

Photometry itself has for decades been performed using apertures. That is, the integrated flux or total brightness of a source is computed within apertures of a fixed size. This is especially useful for isolated, unresolved, point-like sources like stars, quasars, and distant galaxies whose spatial appearance is well-described by the point-spread function (PSF) determined by the optical train of the telescope. While larger apertures ensure all of the light is captured and are less susceptible to noise, they may unintentionally capture light from other nearby sources which is usually mitigated by smaller apertures, although with typically greater uncertainties. Images with high source density, arising either from physically compact structures (e.g., star clusters) or from background and foreground sources appearing in close proximity on the sky, may require apertures smaller than the PSF (or alternative mitigation strategies, see \citealt{Stetson1987, bertin_sextractor:_1996}). Recovering the total flux in such cases requires scaling the aperture-integrated flux proportional to the total extent of the PSF, which often involves complicated strategies to characterize the PSF stability across the detector or co-added mosaic. Transitioning from monochrome photometry of a single band to photometering multi-wavelength images presents its own challenge as PSFs and pixel sizes typically vary with the filter as well as telescope, instrument, and observing conditions. The solution has been a procedure known as PSF homogenization (or matching) whereby each image is convolved with a kernel that maps the PSF of that particular image to that of a target PSF, typically requiring re-sampling images to a common pixel scale. Not only is the choice of the target PSF not always well-defined, especially in cases where the PSF characteristics vary significantly between bands, but re-sampling images often induces or increases pixel-to-pixel covariance.

For applications in extragalactic studies, the deepest wide-field ground-based near-infrared survey at the time of writing is UltraVISTA \citep{mccracken_ultravista_2012} which at a uniform $K_s\approx26$\,AB depth captures $\sim150$ sources per arcmin$^2$ over 2\,deg$^{2}$ with resolution set by its $\sim$0.51\arcsec~PSF at FWHM. Consequently, modest apertures of 3\arcsec\ diameter can be contaminated by neighboring sources. In the corresponding source catalog of \citet{Weaver2022_catalog}, 2\arcsec\ diameter apertures are adopted when measuring photometry to be used in spectral fitting, which in the case of some high-redshift ($z\gtrsim7$) galaxies remain contaminated such that interloping blue light does not permit a high-redshift solution \citep{Kauffmann2022}. While manually removing such interlopers in small samples is possible, doing so for several thousand becomes impractical, and risks imposing human biases. Until the operation of space observatories such as \Euclid{} and \textit{Roman}, surveys with the large area and near-infrared bands necessary to detect large numbers of rare, high-redshift galaxies will continue to be conducted by ground-based facilities at significantly lower spatial resolution, and so these challenges to aperture-based methods will only become more difficult, e.g. 40\% of \textit{Rubin}/LSST sources will be blended at above a 5\% level \citep[see Fig.~19 of][]{Faisst_oddballs}. As we will also demonstrate, the success of aperture photometry becomes more limited with deeper surveys of crowded galaxy fields.

These challenges must be met with appropriate solutions now if we are to continue exploring not only the high-redshift Universe, but pursuing any study whose success relies upon contending with crowded fields and faint sources (e.g., cosmology, transients). Successfully approaches will necessarily be robust to contamination from neighboring sources, provide reliable limits on non-detections, and be consistently applicable over a wide range of spatial resolution, wavelength, seeing, and sensitivity.

An attractive class of alternative photometric techniques called ``profile-fitting'' photometry has enjoyed great success overcoming these very challenges. They work by fitting a model (parametric or non-parametric) that describes and can be reliably fit to the surface brightness profile of a source. Usually the total brightness is a parameter of the model, or can somehow be derived from it. Commonly used parametric implementations of profile-fitting involve a source model parameterized by flux, position, and for resolved sources also size, axis ratio, position angle, and light profile (i.e. S\'{e}rsic index; \citealt{1963BAAA....6...41S}) which is then convolved with a known PSF and fit to the surface brightness profile a given source. This approach has significant advantages over traditional apertures. Firstly, the flux reported is the total brightness of the source in that particular band, avoiding aperture corrections and related systematics. Secondly, the PSF is a property of the model, which is a more tractable solution compared to PSF homogenisation which manipulates the measurement image. This means that the fitted properties of resolved sources are the intrinsic, PSF-deconvolved values. Thirdly, positions are not simply determined as the peak or centroid of an image but are rather fitted parameters, subsequently achieving greater precision over commonly-used peak-finding routines in photometry software (e.g., \texttt{Source Extractor}; \citealt{bertin_sextractor:_1996}). Lastly, sources that have some fraction of their flux overlapping can be accurately photometered by fitting an appropriate number of simultaneous models. This forward-modeling ability to de-blend sources is unique to profile-fitting photometry and means that sources easily differentiated in high-resolution images can be accurately photometered in low-resolution bands such as \textit{Spitzer}/IRAC.

\tractor{}\footnote{\url{https://github.com/dstndstn/tractor}} \citep{lang_tractor_2016} is one such profile-fitting tool. Given a set of initial positions, model profiles (e.g. point-like versus resolved), and image information with per-pixel uncertainties, \tractor{} optimizes those models for a given set of images whose sources have been already identified from some existing (ideally higher resolution) detection image. The key distinction when utilizing such parametric models is that we can derive a likelihood for the particular model parametrization given the data, as well as estimate uncertainties on those parameters. Key implementations of \tractor{} include \citet{2016AJ....151...36L}, \citet{Faisst_oddballs}, and \citet{Stevans_SHELA}. In addition, \citet{Nyland17_servs_tractor} explored for the first time the capabilities of \tractor{} to photometer highly blended IRAC sources using models derived from higher resolution VISTA imaging. 

We develop a pipeline to perform reproducible profile-fitting photometry built around \tractor{} called \farmer{}, which adopts similar principles used in previous work concerning model-based photometry including \texttt{HSCPipe} \citep{aihara_second_2019}, the \texttt{DECaLS} pipeline \citep{dey2019_decals}, \texttt{GaLight} \citep{Ding_galight}, and \texttt{SExtractor++}, \citep{Bertin2020, Kummel2020}. \farmer{} provides a larger framework within which \tractor{} can be scaled to large galaxy surveys where source detection must be handled in a statistical manner. Crucially, \farmer{} includes built-in parallelization methods which enable efficient computational runtimes. \farmer{} utilizes the optimization routines already provided by \tractor{} to obtain estimates of source flux and positions, as well as galaxy shapes for resolved sources. At no point are fluxes derived through integration over an aperture. Instead, the fluxes are derived directly from the normal
ization factor required to scale a unit-normalized model to best describe a given source. Parameter uncertainties, including flux, are derived as minimum-variance estimates according to the Cram\'er-Rao bound \citep{Cramer1946, Rao1945}. For point-like sources, this equates to the classical variance derived when fitting a pattern using inverse-variance weights.

We begin in Section~\ref{sec:review} with a review of the key aspects of \tractor{}. Section~\ref{sec:farmer} then describes the purpose and design of \farmer{}. Section~\ref{sec:validation} presents the results of benchmarking \farmer{} against a set of simulated COSMOS-like images before concluding in Section~\ref{sec:summary}. Other considerations and discussion is included in Appendix~\ref{sec:assumptions}.

The features, behaviour, and performance of \farmer{} described in this paper is purposefully consistent with its use in \citet{Weaver2022_catalog} so that it can be used as a supporting reference. The software is available on GitHub\footnote{\url{https://github.com/astroweaver/the_farmer}} and is provided `as is'. The authors reserve the right to update the software -- and its features and performance -- at any time, documenting relevant changes. The material presented here is independent of any assumed cosmology. All magnitudes are expressed in the AB system \citep{oke_absolute_1974}, for which a flux $f_\nu$ in $\mu$Jy ($10^{-29}$~erg~cm$^{-2}$s$^{-1}$Hz$^{-1}$) corresponds to AB$_\nu=23.9-2.5\,\log_{10}(f_\nu/\mu{\mathrm{Jy}})$.

\section{Review of the Tractor methodology}
\label{sec:review}

\tractor{} is a recent development aimed at providing a generalized framework for fitting the surface brightness profiles of sources in an image. The approach is generative, that is, \tractor{} attempts to construct a predictive model based on the science image, a corresponding PSF, and a per-pixel noise estimate (typically a weight map), and optionally a background sky model; as well as initial guesses as to the model parameters such as source positions, shapes, and fluxes. In practice, \tractor{} optimizes these initial parameters to produce a model image which describes input image within the bounds of the properties provided, separating the source signal from the background noise. 

The flux of a given source $\alpha$ is not measured with apertures, but is rather obtained directly as the normalization of a unit-normalized model profile $G(\phi)_i$, where $\phi$ is the subset of parameters describing the position and shape of the overall model $m(\theta)_i$ defined over every pixel $i$ and convolved with the PSF:

\begin{equation}
m(\theta)_i = \alpha\,G(\phi)_i \circledast \mathrm{PSF_i}
\label{equ:model}
\end{equation}

\noindent The flux for a single isolated point source is essentially computed as a mean of the input image $x_i$ and the model image $m(\theta)_i $ normalized to unity and inversely weighted by pixel variance $\sigma_i^2$. In other words, flux is the value required to scale a unit-normalized model image of a point source to describe the real point source. \tractor{} attempts to maximize the likelihood $\likelihood(x | \theta)$ of the data $x$ given the free parameters $\theta$, and uses the quadrature addition of the weighted residual image (i.e. $\chi$), which is analogous to a $\chi^2$ minimization as $\mathcal{L}\sim\exp(-\frac{1}{2}\chi^2)$ but in two spatial dimensions, ignoring pixel-pixel covariances: 

\begin{equation}
\likelihood(x | \theta) = \exp\left(-\sum_i \frac{(m(\theta)_i - x_i)^2}{2 \sigma_i^2} \right)
\end{equation}

One immediate advantage of this approach is that it avoids the need for PSF homogenization as the PSF is included in convolution with the source profile. Another advantage is that as long as the model is normalized to unity including the wings, it may be truncated in numerical processing without biasing the estimated flux. Therefore while an aperture over the model realized in some restricted image dimensions will return a flux less than the true flux, the flux determined by scaling the unit normalized (but truncated) model will remain accurate. This is especially useful when considering numerical and computational limitations.

Of perhaps equal importance are parameter uncertainties. The uncertainty estimates produced by \tractor{}, reproduced here based on documentation provided with the code, are related
to the Cram\'er--Rao bound which is a lower bound on the variance of any unbiased estimator $\hat{\theta}$:
\begin{equation}
\var(\hat{\theta}) \ge \displaystyle\frac{1}{I(\theta)}
\end{equation}
where $I(\theta)$ is the Fisher Information,
\begin{align}
  I(\theta) &= E_{x}\left[ \left( \frac{\dd \log \likelihood(x | \theta)}{\dd \theta} \right)^2 \right] \\
  & = - E_{x}\left[ \frac{\dd^2 \log \likelihood(x | \theta)}{\dd \theta^2} \right].
\end{align}

\noindent The log likelihood is therefore
\begin{equation}
\log \likelihood(x | \theta) = -\sum_i \frac{(m(\theta)_i - x_i)^2}{2 \sigma_i^2}
\end{equation}
with first derivative
\begin{equation}
\frac{\dd}{\dd \theta} \log \likelihood(x | \theta) = - \sum_i \frac{m(\theta)_i - x_i}{\sigma_i} \frac{1}{\sigma_i} \frac{\dd}{\dd \theta} m(\theta)_i
\quad ,
\end{equation}
which should equal zero when the likelihood has been maximized.

The second derivative is

\begin{multline}
\frac{\dd^2}{\dd \theta^2} \log \likelihood(x | \theta) = \\
    - \sum_i \left[\frac{m(\theta)_i - x_i}{\sigma_i} \frac{1}{\sigma_i} \frac{\dd^2}{\dd \theta^2} m(\theta)_i + \left( \frac{1}{\sigma_i} \frac{\dd}{\dd \theta} m(\theta)_i \right)^2 \right]
\quad ,
\end{multline}

\noindent where the first term is zero at the optimum. Returning to the Cram\'er--Rao bound, we have

\begin{align}
\frac{1}{\var(\hat{\theta})} \le 
& - E_{x}\left[ \frac{\dd^2 \log \likelihood(x | \theta)}{\dd \theta^2} \right]
\label{eq:secondderiv}
\end{align}

\noindent and since our second derivative (equation \ref{eq:secondderiv}) is independent of $x$, the expectation
collapses and we get

\begin{align}
\frac{1}{\var(\hat{\theta})} & \le 
  \sum_i \left( \frac{1}{\sigma_i} \frac{\dd}{\dd \theta} m(\theta)_i \right)^2
  \quad ,
\end{align}

\noindent which is the inverse-variance estimate reported by \tractor{}. In the important case of estimating flux where $\theta\equiv\alpha$, the derivative of the model with respect to flux is just the profile of the model.  Hence, the uncertainty estimate on flux for point-like sources is based entirely upon the PSF and the per-pixel error estimates $\sigma_i$ from the weight map. 

We can gain a better understanding of \tractor{}, both its functionality and limitations, through progressively complex examples.

The simplest example is an isolated, point-like galaxy. \tractor{} is supplied with the image, a weight map, a PSF, and a known position for the source; \tractor{} does not provide means to detect sources, and so a list of initial source positions is required beforehand. While the data input (image, weight map, PSF, optionally sky) must be kept fixed, we may also fix the position parameter so that only the flux is allowed to vary. This one parameter optimization is linear in the case of a single source. However, profile-fitting photometry is sensitive to offsets in source positions requiring greater precision than is typically needed for accurate aperture photometry. One can address this by simply allowing the model position to also vary, and \tractor{} has built-in functionality to deal with this. This three parameter (i.e. x, y, flux) optimization is a non-linear procedure, although the degeneracy between the position and flux parameters should be virtually zero. The result is not only an estimate of the flux, but also the source position. The source may also be photometered in many bands in a single joint optimization where the shape and position are shared but flux is now a vector with an element for each band. 

A more complicated example is an isolated, resolved source. \tractor{} includes a library of discrete parametric models which include but are not limited to, in order of simplicity, point source profiles taken from the PSF stamp (as assumed in the previous example), resolved models with exponential or de Vaucouleurs profiles \citep{deVaucouleurs1948}, full S\'ersic profiles, and composite profiles made by superimposing exponential and de Vacouleurs profiles. As before, the properties of the input data (i.e. image pixel values) are kept fixed. We also may fix the position, for simplicity, leaving the source shape and flux free to vary. The question then is how to decide which shape parameterization to use? \tractor{} does not provide an answer; rather it is up to the user to determine a model type ahead of the optimization. A resolved model type is appropriate in this case, and so now our optimization returns source fluxes and shapes (e.g., effective radius, axis ratio, and position angle). Photometry of other images taken with different filters is usually of interest and so by fixing the model shape we can perform ``forced photometry''. Although it is possible to allow the shape to vary with each band, this comes at the cost of potentially overfitting our model.

An even more complicated example is an image containing many sources of various morphological presentations and crowding. This is typically what is encountered in deep galaxy surveys and presents a serious challenge. We have already understood that \tractor{} does not provide source detection, and so the degree to which the photometry succeeds is dependent on the performance of some external detection procedure. Once we have somehow supplied source centroids to \tractor{}, we are still left to determine the appropriate model type for each source. Although it may be feasible to assign model types manually for small regions of interest occupied by a small number of sources, this is typically not practical for large surveys containing thousands or even millions of sources. Assuming this can be done in some way, \tractor{} will optimize all source models simultaneously on that given image to produce optimized shapes which can then be fixed to performed forced photometry on other bands of interest. Alternatively, one can use all the bands of interest to optimize the model and simultaneously obtain measurements of fluxes, although this adds significant complexity that may cause the optimization to not converge.

The situation does not improve much even if there is only one source of interest amongst a crowded field. Although one may try to instantiate a single model at that source position, \tractor{} uses information from every pixel in the image that has non-zero weight. That means that the presence of every other source in the image counts against the likelihood. One option is to restrict the weight map to only the pixels belonging to that source. However, deciding the extent of such a region is non-trivial. Regions that are too large may include flux from a neighbor which are unaccounted for by our one source model, and may bias the photometry typically towards higher fluxes. Having too small an region is suboptimal, and ill-defined as you would need to know the extent any neighbors beforehand. Another option is to continue instantiating models (defined by centroids and model types) for all nearby sources until it is possible to cleanly define a contiguous region whose boundaries do not contain light from other sources (i.e., an isolated group of sources). Such a manual approach may work, but only in limited cases where the user is heavily involved, severely limiting reproducibility. Even if this can be done, it is not immediately obvious how best to fit this potentially large group of nearby sources. Should they be fit simultaneously? This approach is straight-forward but computationally expensive. Perhaps they should be fit one by one, subtracting the best-fit model each time? This is usually computationally faster, but induces hysteresis that can bias photometry.

\begin{figure*}[t]
	\centering
	\includegraphics[width=1\hsize]{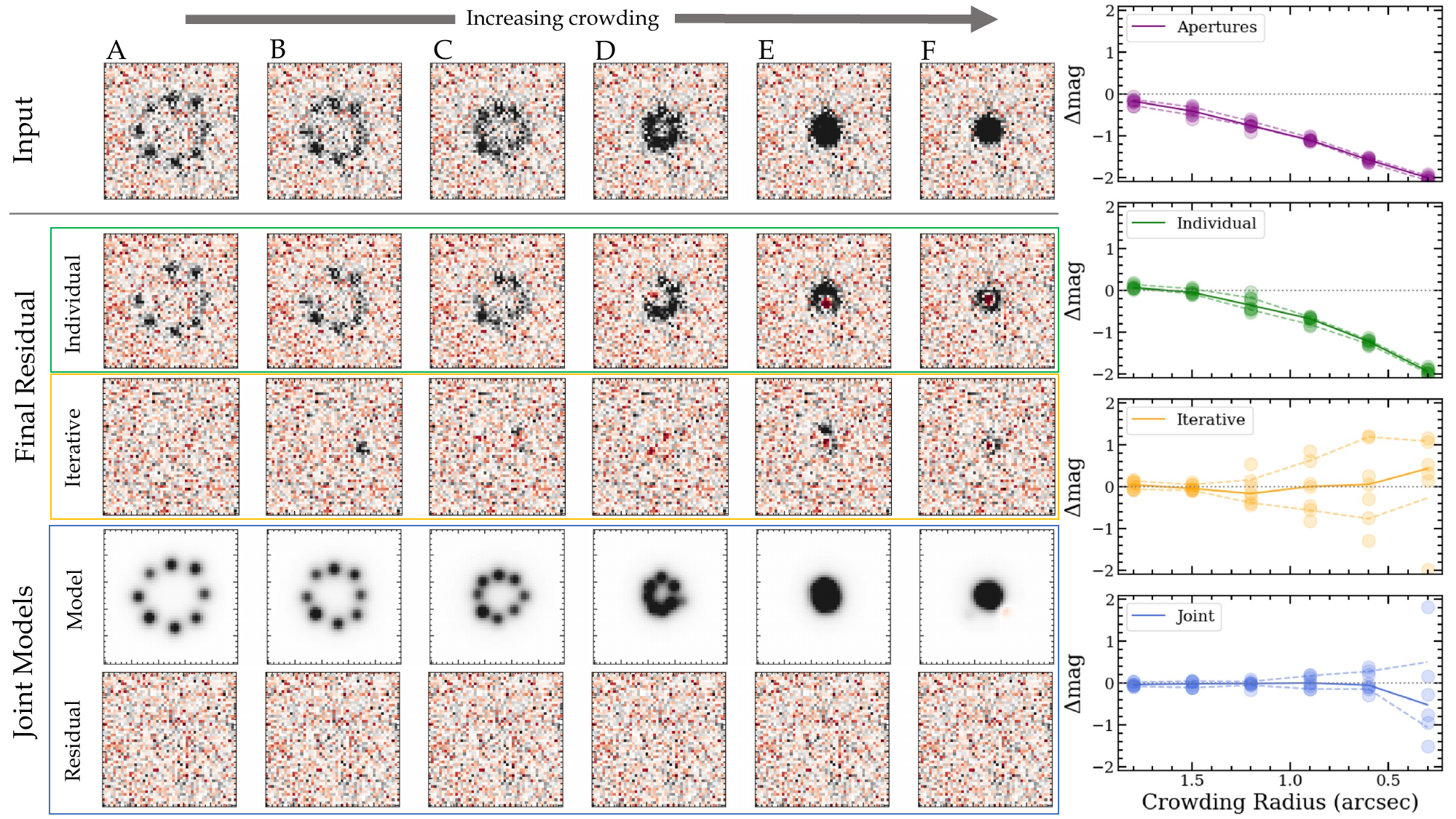}
	\caption{Eight similarly bright point sources are injected into a simulated noise field over six scenarios of increasing degrees of circularly symmetric crowding. Assuming source positions are known beforehand, fluxes are measured in four ways: 2\arcsec apertures, profile-fitting each source independently, profile-fitting each source with successive model subtraction, and jointly fitting all models simultaneously. The degree of success of each method is shown on the right measured in the difference in magnitude $\Delta{\rm mag}$ between the input and measured magnitudes as a function of source crowding, with a median $\Delta{\rm mag}$ and 68\% ranges indicated for each scenario. Only joint fitting provides both precise and accurate recovery of crowded sources. 
	}
	\label{fig:demo_photmethods}
\end{figure*}

A generalized version of this dilemma is useful in proving this point. In Figure~\ref{fig:demo_photmethods}, eight point sources are injected into a Gaussian noise field at signal strengths ranging from $\sim3-10\,\sigma$ and arranged in a circle. A total of six cases are constructed (A, B, C, D, E, F) by varying the radial distance to each source such that at one extreme they are separated (A) and overlapping at the other (F).

As a baseline, fluxes are summed in 2\arcsec\, apertures that do not overlap in case A and so recover accurate fluxes. However, a bias grows towards case F where the apertures become confused and eventually include the flux from all eight sources in each aperture. This highlights the limitations of apertures in pathologically crowded fields, after which one must appeal to statistical mitigation strategies afterwards to re-scale fluxes. We move on to profile-fitting photometry in the subsequent rows. The most direct approach is to model each galaxy individually in series (allowing the uncertain positions to necessarily vary in the fit), but by case C succumbs to the same confusion as the apertures and multiply counts each source per model. An attractive solution is to also iteratively subtract each model one by one, in series. While this approach is certainly more successful in that the average flux is accurate, that of most individual sources is catastrophically wrong. This poor performance is also evidenced by significant residual flux in all but the least crowded case.

The optimal way is to model each source simultaneously, which allows the \textit{joint} model to recognize that there are neighbors that it can describe. This approach does not suffer from the drawbacks of fixed apertures, or of fitting models individually or with subtraction. It recovers unbiased photometry in cases A, B, C and D, failing in only the most crowded cases (E and F). Yet, this level of crowding is pathological as it is unlikely that (in the absence of higher resolution data) a source detection procedure would be able to separate the signal into even two centroids, let alone all eight. Therefore the most extreme cases remain a problem, but one which will have to be addressed by innovations in source detection and associated de-blending techniques. Although fitting multiple nearby sources simultaneously is clearly the optimal approach, it is also the most computationally expensive one, and for that reason it cannot be so readily scaled up to large area surveys without first developing efficient algorithms that can be utilized successfully by high performance computing facilities.

As we can see, \tractor{} is a powerful tool for determining best-fit values corresponding to parametric models of sources, but it requires significant manual attention in all but the simplest cases. Therefore there is a considerable gap between the function of \tractor{} and what is required for front-to-back catalog pipelines. Developing such pipelines is not only time consuming, but independently developed pipelines perform differently (e.g., that of \citealt{Nyland17_servs_tractor} is different than the pipeline of \citealt{dey2019_decals}). While each implementation may be optimized for a certain task, the overwhelming success of software like \texttt{Source Extractor} is that they are immediately accessible, flexible, and easy to use. However, the matters of source detection, model type decisions, which groups of sources to model and how best to model them, as well as computational efficiency are challenges that must be solved if we are to construct such a generalized p
ipeline that applies \tractor{} to the incredibly deep, crowded fields to be explored by the next generation of galaxy surveys.

\section{The Farmer: A general description}
\label{sec:farmer}

\texttt{The Farmer} is a generalized, flexible, and reproducible framework that uses the model library from \tractor{}, its optimization engine, and several helper routines to photometer detected sources, measure their shapes, produce catalogs and ancillary images, as well as provide supporting diagnostics. \texttt{The Farmer} overcomes the issue of how to assign model types by identifying natural groups of nearby sources and determines the best model type of each source using a decision tree in a time efficient, optimal way whilst mitigating related pathological situations. It includes a significant organizational capacity such that images can be divided up into sections for massively parallelized computation. Here we walk though the process of \texttt{The Farmer} from image preparation to the output catalogs.



\subsection{Image Preparation}
\label{sec:image_prep}


At bare minimum, \farmer{} requires a single science image containing sources of interest. A corresponding inverse variance weight map is ideal, but not required. Lacking weight information, \farmer{} can utilize the Pythonic \texttt{Source Extractor} code \texttt{SEP} by \citet{barbary_sep_2016} to measure noise directly from the images or simpy assume equal weights. 

In this basic case, \farmer{} will detect sources, model them, and perform forced photometry all on the same monochromatic image. In more typical, complex cases it is desirable to produce a separate detection image that combines multiple bands. For surveys of faint sources, the \texttt{CHI-MEAN} approach \citep{Szalay99_chidetection, Bertin2010_swarp} has been widely adopted \citep[e.g.][]{laigle_cosmos2015_2016, Weaver2022_catalog}, or a similar signal-to-noise image co-add (e.g. \citealt{Whitaker2011}).

Masking is especially important in profile-fitting photometry for the reason that it is inadvisable to attempt to model large, saturated stars, nebulae, or nearby galaxies which are essentially nuisance foreground contamination. While apertures have the advantage of being able to efficiently sum fluxes in whatever regions of an image are of interest, models must attempt to describe the image as it is. Attempting to model such nuisance sources, which lie outside the reach of our parametric models, will never achieve a satisfactory fit even after several hundred central processing unit (CPU) hours, if at all. That being said, we note without extensive sky background modelling, sources within bright star halos (e.g.) will not be photometered accurately with apertures either. 

A useful recipe is to stack all bands that will used to detect sources, and mask out the full extent of such nuisance foreground objects, and possibly also the edges of the mosaic or detector. \farmer{} can be configured to apply a mask before or after source detection. The latter is preferred in virtually all cases, as mask edges can produce spurious sources. Applying a mask after source detection simply removes sources from the catalog and their corresponding segments are zeroed out.

\farmer{} includes several ways to measure image backgrounds and per-pixel noise based on \texttt{SEP}, and this can be configured by the user. Backgrounds can be measured as global medians or spatially varying (following the methods of \texttt{Source Extractor}; see \citealt{bertin_sextractor:_1996}), with per-pixel noise being estimated directly from the RMS of the image. The background and per-pixel noise estimates can be produced with and without the mask in order to mitigate the adverse effects of bright stars and foreground galaxies. Although currently all detected sources will be modelled, the ability to identify and remove spurious sources on-the-fly is expected to be included in a future update.

\subsection{PSF creation}
\label{sec:psf_creation}

With the images and weights in hand, \farmer{} needs a PSF for each band of interest. There are many way of generating PSF stamps, including as realizations of spatially varying models, and \farmer{} can be supplied with several PSF types. 

The most common is a constant PSF stamp sampled at the same pixel scale as the its corresponding image; these can be readily produced by packages such as \texttt{PSFEx} \citep{bertin_psfex_2013}. One may also use \texttt{PSFEx} to generate spatially varying PSF models, all flavours of which (e.g., Gauss-Laugerre or pixel bases) are understood by \farmer{} (and importantly also by \tractor{}).  While this can be achieved through using \texttt{PSFEx} by itself, \farmer{} is able to run \texttt{PSFEx} in a semi-automatic way using built-in functions. First \farmer{} runs \texttt{Source Extractor} to identify bright sources and produce a catalog including vignette stamps that \texttt{PSFEx} can read in (\texttt{SEP} do not produce such output files). Candidate point sources are then selected either automatically by \texttt{PSFEx}, or more directly by a pre-selection by the user based on source FWHM and brightness. The user can also declare which bands should use a constant PSF and which should be spatially varying, and \farmer{} will automatically reconfigure \texttt{PSFEx} in each case. 

In some cases the PSF varies too quickly across an image to be accurately characterized by a smoothly varying surface as used by \texttt{PSFEx}. It is possible therefore for the user to supply a set of PSFs and a file which maps each one to a coordinate so that \farmer{} can use the nearest sampled PSF for a given source. The assumption of a smoothly varying PSF can thereby be avoided, and the user is free to choose the grid geometry according to their requirements. This `PSF Grid' approached was developed in \citet{Weaver2022_catalog} to characterize the photometry of the Subaru Suprime-Cam mosaics in COSMOS.

The images of \textit{Spitzer}'s Infrared Array Camera (IRAC) feature a highly variable PSF which is generally triangular in shape. The \texttt{PRFMap} package\footnote{\url{https://github.com/cosmic-dawn/prfmap}.}) attempts to characterize this highly irregular PSF by mapping the pixel of each stacked image back to the locations on the CCD of the constituent images. It then uses the spatially-dependent calibration PSFs to construct a combined PSF for the stacked image. Similar to the PSF Grid approach, \texttt{PRFMap} produces a library of individual PSFs corresponding to a fixed grid of sampling coordinates. This output can be used with \farmer{} to measure IRAC photometry.

One important caveat to note is that in all cases the PSF must be measured into its wings and not be truncated. This is for two reasons. Firstly, profile-fitting models generally benefit from the wings of the PSF being in tact. This can be immediately appreciated in the case of unresolved sources fit with point-source models for which \tractor{} uses the PSF stamp for the model profile: if the wings of the point-source model do not describe the full spatial profile of the source of interest (i.e. all pixels within the source segment that contribute to its $\chi^2$) then the residual will always have leftover signal in the wings and the measured flux may be biased. Secondly, the pixel values of a PSF which has been truncated and then normalized to unity will be larger than those of the full PSF normalized to unity, and so its optimal normalization coefficient (i.e. flux) will be smaller for the same source, introducing a bias. Therefore it is strongly advised to sample the entire PSF profile out to radii where the wings are indistinguishable from noise, in most cases corresponding to a radius of several arcseconds.

\subsection{Source Detection}
\label{sec:detection}

The first step in catalog creation is source detection. \farmer{} utilizes \texttt{SEP} \citep{barbary_sep_2016} to provide source detection, segmentation maps, background, and noise estimation with near identical performance as classical \texttt{Source Extractor}. As with any detection software, the performance of \texttt{SEP} as measured by source de-blending and segmentation, e.g., is entirely dependent on the configuration of the detection parameters set by the user (see \citealt{bertin_sextractor:_1996, Szalay99_chidetection, Holwerda2005}). Segmentation of blended sources in both \texttt{SEP} and \texttt{SourceExtractor} relies on multiple thresholds to determine which pixels belong to which galaxy (see Section~2.3.1 of \citealt{Haigh2021} for details). Generally, detection strategies vary between catalogs, are typically driven by science objectives, and are often tuned by eye. The performance of \farmer{} described in this work is no different; for the sake of comparison to COSMOS2020 we adopt the same detection configuration as described in \citet{Weaver2022_catalog}.

We stress that although SEP is responsible for identifying individual galaxies, the deblending of their light is entirely determined by \farmer{}. However, if SEP incorrectly groups separate galaxies together, \farmer{} cannot deblend them afterwards (see Section~\ref{sec:deblend}). Detection parameters for \texttt{SEP} can be configured directly with \farmer{}, and related diagnostic images are supplied indicating source centroids on the detection image. It is also possible to hand \farmer{} a catalog of source coordinates and a corresponding segmentation map from e.g., \texttt{Source Extractor}, or any other similar detection software.

\begin{figure}[t]
	\centering
	\includegraphics[width=1\hsize]{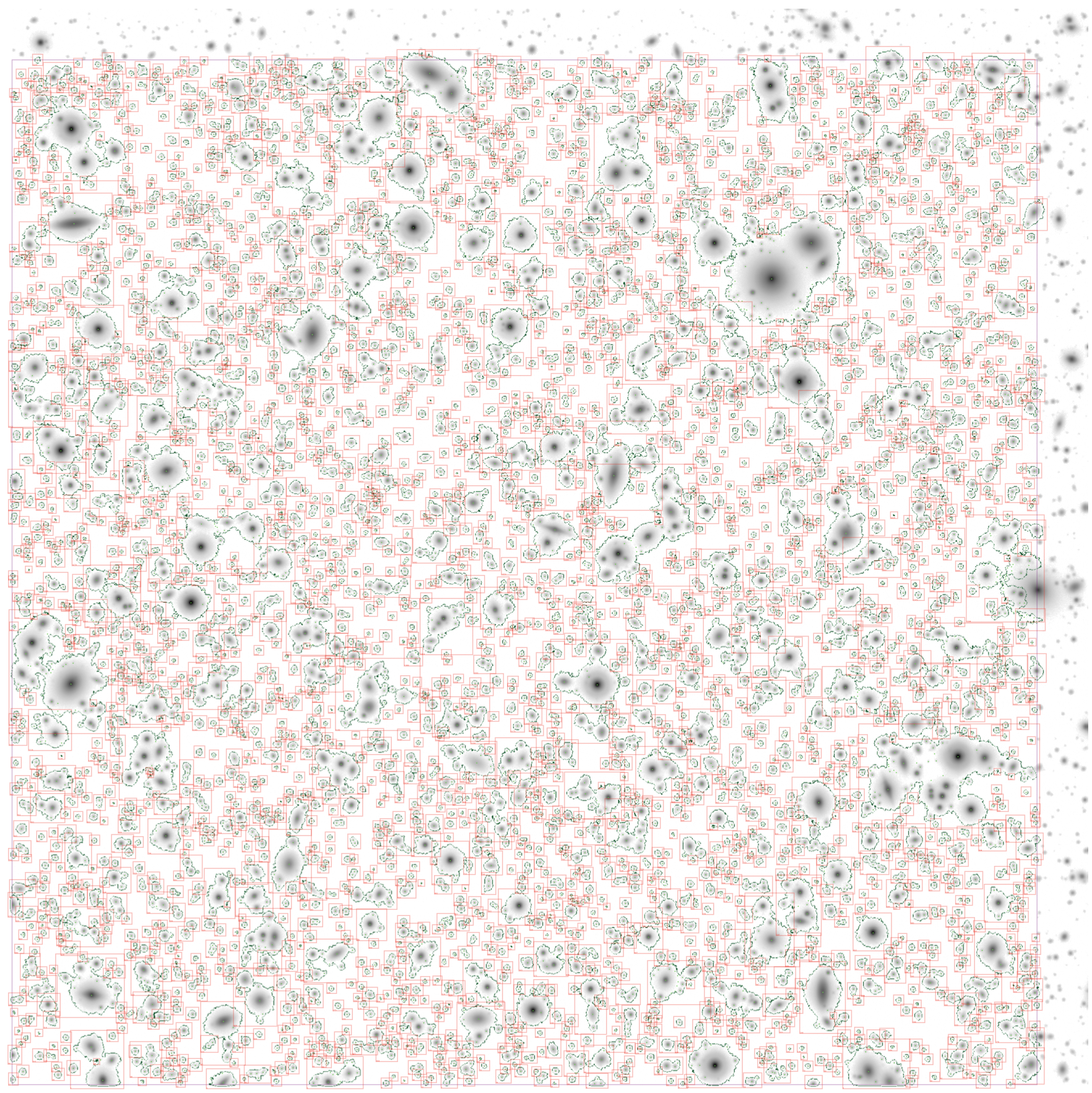}
	\caption{Example of groups detected over a brick. This brick lies at the edge of a mosaic and so has boundaries with two adjacent bricks. Source centroids are shown by green crosses. Groups are outlined by red boxes and group footprints of owned pixels by the green borders.
	}
	\label{fig:blobmaster}
\end{figure}

\farmer{} performs all functions on discrete sections of the total mosaic called ``bricks'' (following \citealt{dey2019_decals}). An example is shown in Figure~\ref{fig:blobmaster}. Each brick is cut out of the total mosaic image, weight, and mask with equal dimensions, and includes a overlap region on each side. Sources detected with centroids in the overlap region are removed from the source catalog of the brick, and the pixels belonging to their segment are set to zero (like background pixels). They are not lost, however, as they are found again in the main region of a neighboring brick. This ``fuzzy boundary'' approach means \farmer{} can construct unique source catalogs for each brick which have no overlap with neighboring bricks, thus accounting for every source without duplication or loss. Although the overlap regions of the segment map are also set to zero, \farmer{} keeps segment pixels in the overlap region of sources whose centroids are in the main region of the brick. This behavior allows sources which are near the overlap zone to be modelled with all of their pixels, as opposed to a strict cut-off at the overlap boundary where their profiles would be truncated.

Following the creation of the brick's preliminary source catalog provided by \texttt{SEP} and a cleaning of the overlap regions, \farmer{} attempts to identify natural groups of detected sources which would benefit from being simultaneously modelled. Groups are identified by dilating the original segments to form contiguous non-zero regions. Sources which are not in crowded areas form singularly occupied groups, whereas sources in crowded regions end up members of larger groups to be modelled simultaneously. See Section~\ref{sec:limit_detection} for further discussions.

\subsection{Model Type and Shape Determination}
\label{sec:make_models}

A model must now be determined for each source in a given group. The goal is to not only determine the most suitable model for each source, but also its best-fit parameters. While the number of possible decision tree architectures is virtually infinite, \farmer{} relies on a balanced architecture consisting of five discrete models to describe resolved and unresolved, stellar and extragalactic sources:

\begin{enumerate}
    \item \textbf{PointSource} models are taken directly from the PSF used. They are parameterized by flux and centroid position and are appropriate for unresolved sources.
    \item \textbf{SimpleGalaxy}\footnote{SimpleGalaxy models are not a standard \tractor{} model, see \url{https://github.com/legacysurvey/legacypipe}.} models use an exponential light profile with a fixed user-defined effective radius such that they describe marginally resolved sources and mediate the choice between PointSource and a resolved galaxy model. They are parameterized also by flux and centroid position.
    \item \textbf{ExpGalaxy} models use an exponential light profile. They are parameterized by flux, centroid position, effective radius, axis ratio, and position angle.
    \item \textbf{DevGalaxy} models use a de Vaucouleurs light profile. They are parameterized by flux, centroid position, effective radius, axis ratio, and position angle.
    \item \textbf{CompositeGalaxy} models use a combination of ExpGalaxy and DevGalaxy models. They are concentric, and hence share one centroid. There is a total flux parameter as well as a paramter for the fraction of total flux assigned to the DevGalaxy component\footnote{CompositeGalaxy models assume the \texttt{FixedCompositeGalaxy} model class in \tractor{}.}. Each component has their own effective radius, axis ratio, and position angle.
\end{enumerate}

In practice, these spatially-resolved models are optimized using sigmoid-softened ellipticities provided by \tractor{} (i.e. the \texttt{EllipseESoft} class), which allows for an unbounded parameter space more suitable for numerical computation. Also note that although ExpGalaxy and DevGalaxy models can be generalized under a single S\'eric model with variable S\'ersic index, we purposely forgo this additional free parameter as it is generally under-constrained by the relatively low resolution imaging of COSMOS2020.

\begin{figure*}[t]
	\centering
	\includegraphics[width=1\hsize]{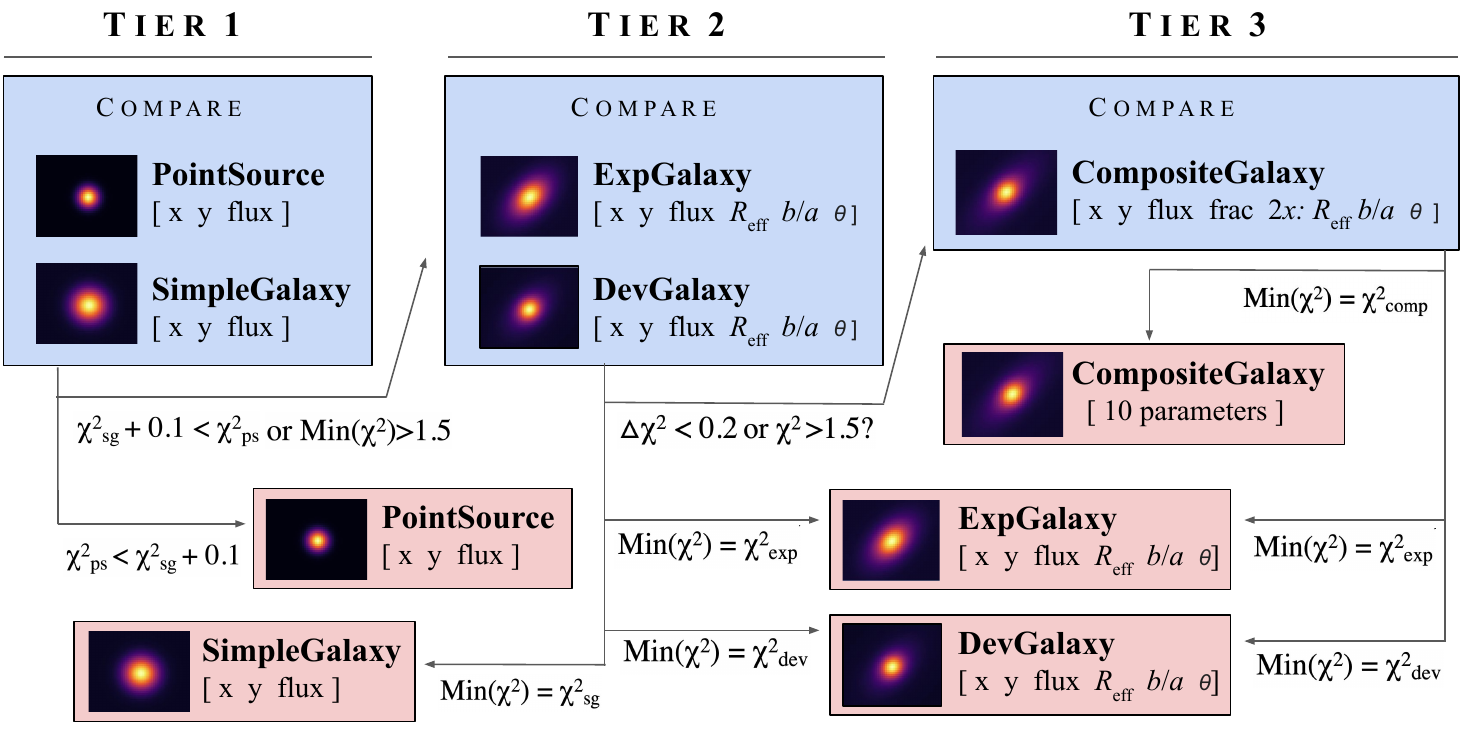}
	\caption{Schematic of the three-tiered decision tree used by \farmer{} to determine the most suitable model type for a given source. The five models are tested in order of increasing number of parameters. Values shown are examples, and should be configured by the user to suit their dataset.
	}
	\label{fig:farmer_decision}
\end{figure*}

\begin{figure*}[t]
	\centering
	\includegraphics[width=1\hsize]{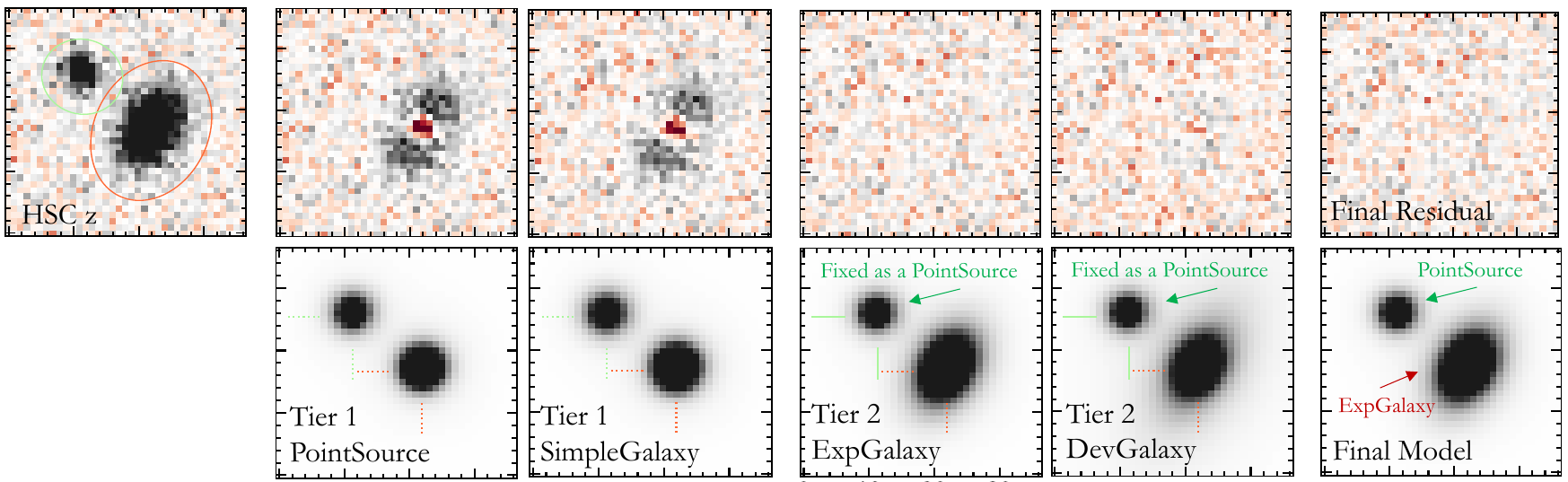}
	\caption{Example of a decision tree process for a group containing two identified sources. The input $z$-band image is shown in the top left with colored ellipses around the two detected objects beside the final residual constructed from subtracting the best model determined by \farmer{}. Shown rightward are the residuals and models for two of the three tiers in the decision tree from which the best model types were determined. This particular pair of sources satisfied the decision tree before reaching the CompositeGalaxy in tier 3. While model images are scaled by $\mathrm{log}_{10}$ to highlight morphologies, the science image and residuals are scaled to $\pm 3\sigma$ to highlight faint signal and any oversubtraction. 
	}
	\label{fig:dt_example}
\end{figure*}

These five models form \farmer{}'s decision tree, whose goal is to both determine the most suitable model for a given source, and provide an optimized set of parameters to describe the shape and position of that source. To ensure that crowded regions do not suffer from poor modelling as a result of the model of a particular source being constrained by light from neighboring source, the models are determined simultaneously at each stage of the decision tree. The values for the decision tree parameters quoted here are examples taken from \citet{Weaver2022_catalog} but can and should be tuned by the user for other data sets. An example of a group containing two sources progressing through the decision tree is shown in Figure~\ref{fig:dt_example}. 

\tractor{} uses a likelihood cost function to score the performance of the joint model containing all of the individual models of the sources in a group. All weight pixels outside the group footprint are set to zero (i.e. no weight) such that nearby sources which are not part of the group cannot influence the likelihood. However, we also need to be able to assess the performance of an individual model for a given source in our group. \farmer{} adopts $\chi^{2}_{N}$ as its goodness of fit statistic, which is calculated by quadrature addition of the residual image pixels belonging to a particular source by its original segment and then reduced by dividing by the number of degrees of freedom $N$, taken as the difference between the number of pixels in its segment and the number of free parameters. 

\farmer{} begins by considering PointSource models for every source in a group, using centroids and fluxes estimated by \texttt{SEP} as initial conditions. \tractor{} then performs an optimization to maximize the combined likelihood of the entire joint model, after which \farmer{} computes the $\chi^{2}_{N}$ for each source in the group. Next, SimpleGalaxy models are considered for all sources in a group with the same initial conditions as before. The models are optimized and the $\chi^2_N$ per source is computed considering pixels within each segment. \farmer{} then tries to place each source into one of three categories: either the source is well fit by the PointSource and is fixed as a PointSource, it is fit well by a SimpleGalaxy, or neither model is appropriate. Satisfying either of the last two categories advances the source down the decision tree towards more complex, resolved model types. The role of the SimpleGalaxy here is not to be a commonly chosen model, but rather a fast-to-compute indicator of a resolved source. Unlike comparing a PointSource model to a more complex ExpGalaxy model, the comparison with the SimpleGalaxy is not only computationally faster but is statistically fair since the number of parameters for both PointSource and SimpleGalaxy are the same, as are the number of data points. Sources which are best fit with PointSource models will be assigned a PointSource model thereafter, which in the case of a one source ``group'' will conclude the decision tree. A source that is better fit by a SimpleGalaxy model by only a slim margin is typically sufficiently modelled by a PointSource also. It is desirable therefore to prefer a PointSource in these cases as a better fit. However, a source well fit by a SimpleGalaxy model triggers the more complex tiers of the decision tree, meaning that the overall group model becomes more complex which requires even longer computational times. \farmer{} therefore penalizes the SimpleGalaxy models in $\chi^{2}_N$ by 0.1 such that a SimpleGalaxy model must have a lower $\chi^{2}_N$ by a margin of 0.1 or better in order to not choose a PointSource (these values again being examples suitable for the COSMOS2020 catalog as empirically determined in \citealt{Weaver2022_catalog}). A PointSource will also not be chosen (at this stage) if produces a bad fit, assessed by $\chi^{2}_N>1.5$, whereupon the source continues to the next level of the decision tree. However, a PointSource or SimpleGalaxy may still be chosen in the end, but only if it is still favored after the assessment of more complex models.

The next stage of the decision tree determines the general Sérsic light profile of resolved sources whose model types remain unfixed, choosing between ExpGalaxy or DevGalaxy. At this stage, fixed sources can only have been assigned PointSource models. \farmer{} starts by considering ExpGalaxy models for all other unfixed sources, performs the optimization, and determines $\chi^{2}_N$ for each. Initial guesses for shape parameters are initialized borrowing from \texttt{SEP} measurements (e.g. $a$, $b$, $\theta$) estimated at detection. Then \farmer{} performs the same computation but with DevGalaxy models on all unfixed sources. Again the $\chi^{2}_N$ is a fair comparison as the number of degrees of freedom are identical between the two model types. \farmer{} allows the model parameters to remain variable for all sources, regardless of whether they have been assigned a final model type, at each stage of the decision tree (e.g., fixed PointSource models still re-optimize their flux). Sources whose ExpGalaxy and DevGalaxy models both fail to achieve a lower $\chi^{2}$ than the SimpleGalaxy are fixed as SimpleGalaxy models, unless the SimpleGalaxy also fails to achieve a $\chi^{2}_N$ of 1.5 in which case that source advances down the decision tree to the third tier. The choice between ExpGalaxy and DevGalaxy models is determined by the lowest $\chi^{2}_N$, without any penalties. However, if the absolute difference in $\chi^{2}_N$ between the two models is less than 0.2, or neither ExpGalaxy or DevGalaxy achieves a $\chi^{2}_N$ of 1.5, the source also advances to the third tier.

All sources have typically been assigned a fixed model by this stage, especially those that have smooth light profiles or are unresolved, and the decision tree ends without trying more complex, time intensive models that \farmer{} has already determined are not required for a sufficient fit. However, highly spatially resolved sources that have reached the third tier without an assigned model are fit assuming the most complex CompositeGalaxy models. If the CompositeGalaxy model fails to achieve a better $\chi^{2}_N$ than either ExpGalaxy or DevGalaxy, the source is assigned the model type that achieved the lowest $\chi^{2}_N$ overall. 

Now that models for all sources belonging to a given group are assigned, \farmer{} optimizes the models a final time. This is an important step as it is possible for an otherwise pathological case to arise whereby two assigned models were never optimized at the same time and their fits may influence each other. By computing this final optimization, the overall likelihood of the model set for the group of sources tends to improve.

\subsubsection{Forced Photometry}
\label{sec:forced_phot}

Now that models types have been assigned and their parameters optimized for each source in a given group, it is straight forward to apply these parametric models to photometer the sources in other bands of interest. We can do this via ``forced'' photometry is to measure fluxes and their uncertainties for already known (detected) sources, fixing the model shape parameters and only allow flux ($\alpha$ in Equation~\ref{equ:model}) to vary. However, \tractor{} provides the flexibility to allow shapes and positions to vary as well; they can be unbounded or limited by a Gaussian prior. For example, it may be desirable to allow the shape to change in the presence of morphological differences between the model bands and the forced photomety band, or allow the position to vary if there are significant astrometric offsets (see discussions in Section~\ref{sec:limit_models}). \farmer{} enables the user to choose which parameters (if any) are fixed during the forced photometry stage.

\begin{figure*}[t]
	\centering
	\includegraphics[width=1\hsize]{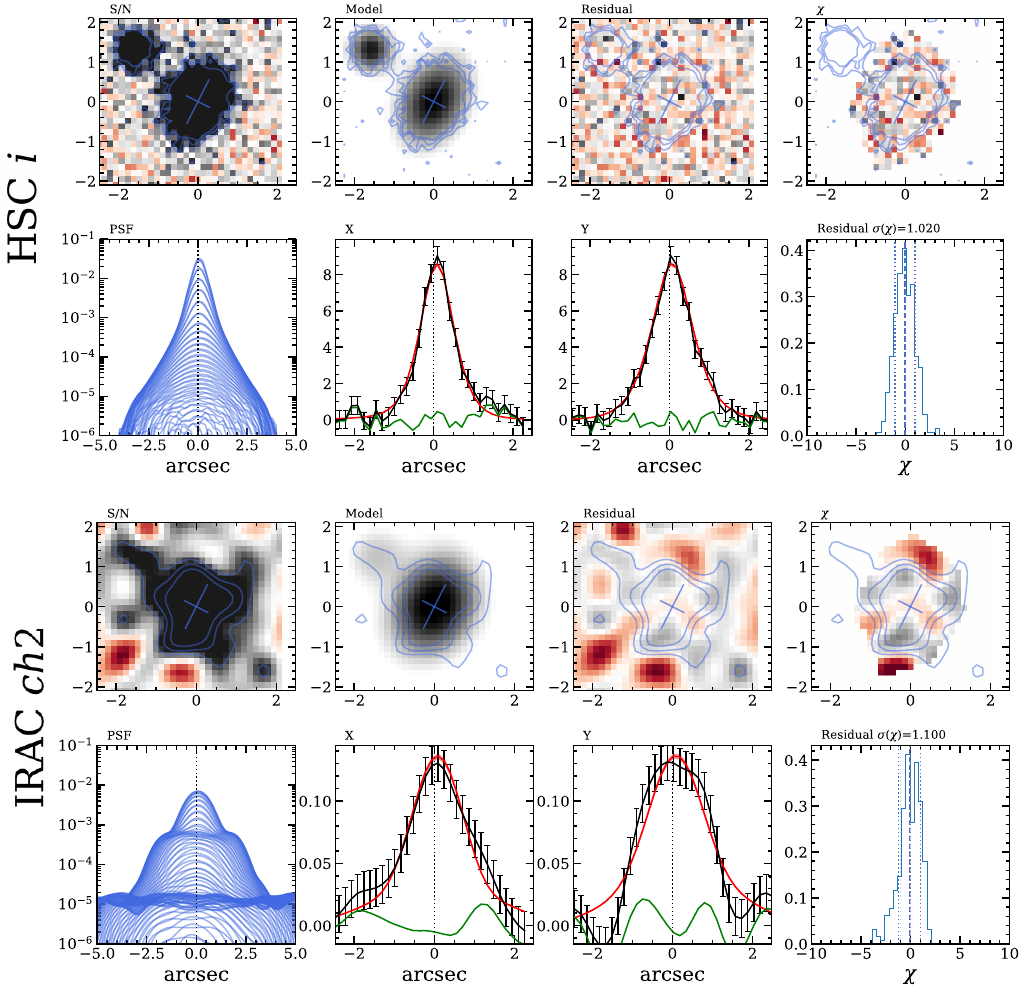}
	\caption{Example of results from forced photometry for the brightest source in Figure~\ref{fig:dt_example} measured in $i$ and channel~2. The image of each band is shown next to the best-fit model from \farmer{}. Brightness contours and principal axes are overlaid on the model in blue. The residuals are shown $\pm\,3\sigma$ (same as the image)  to highlight faint signal and any over subtraction. The rightmost panel shows the per-pixel $\chi$ image scaled $\pm\,3$ computed within the bounds of source segment Bottom rows for each band show a wiremesh representation of the PSF profile; slices though the source (black), best-fit model (red), and residual (green); and the distribution of $\chi$ values over the group pixels which on expectation should be normally distributed. The median and 68\% range of the distribution is shown for illustration.
	}
	\label{fig:fphot_example}
\end{figure*}

As before, fitting proceeds on a group-by-group basis so that the forced photometry can benefit from the same advantage as in the model stage by simultaneously optimizing all models belonging to a given group. Each model is convolved with the PSF of the band of interest and realized into the frame of the image, including images of different pixel scales to that of the detection image\footnote{Currently, \farmer{} requires pixel scale homogenization, but this restriction will be removed in a future update.}. The group models are then simultaneously optimized until their joint likelihood converges, or until some maximum iteration set by the user. Figure~\ref{fig:fphot_example} shows the results of forced photometry using the same sources from Figure~\ref{fig:dt_example}. While this procedure is generally faster than the model stage, forcing photometry on dozens of images may approach a similar computational expense. For consistency, it is advisable to perform forced photometry for all bands, even if they were used in the modelling stage. Computational strategies are discussed in Section~\ref{sec:computation}. 

\subsubsection{Catalogs and Other Output}
\label{sec:catalogs}

After the modeling stage, \farmer{} produces an intermediate catalog containing the source IDs, including their brick and group numbers, followed by the detection parameters from \texttt{SEP}. For each source, the best-fit model type (e.g., PointSource or ExpGalaxy) are recorded, as well as their best-fit parameters and associated uncertainties. Shapes and sizes are not measured for sources assigned unresolved models (e.g. PointSource and SimpleGalaxy). Fluxes and flux uncertainties are also measured for each source in every band used in the modelling stage. 

\begin{figure*}[t]
	\centering
	\includegraphics[width=1\hsize]{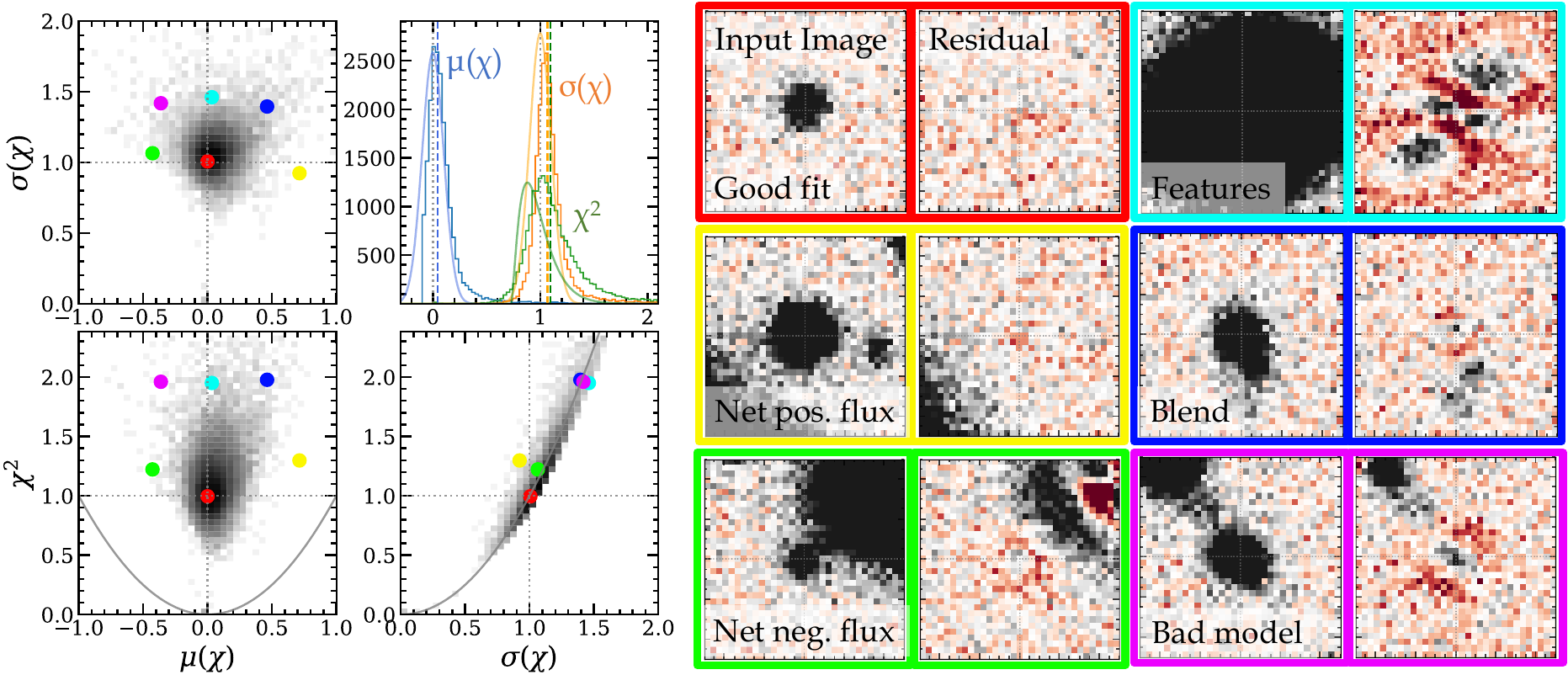}
	\caption{\farmer{} provides detailed statistics to readily select problematic sources such as blends, poor model fits, and artifacts that are difficult to identify from $\chi^2$ alone. The red, yellow, and green sources are well-fit with a small variation in negative and positive flux. The cyan source contains features not well described by our smooth model profiles, the blue source is an unidentified blend, and the purple source is a pointsource model assigned to a resolved galaxy. Each of these six sources is selected from unique regions of a 3D statistic space (left). Each statistic is measured within the residual segment pixels of each source and include the $\chi^2$, the median $\chi$ distribution $\mu(\chi)$, and the standard deviation of the $\chi$ distribution $\sigma(\chi)$. $\chi^2$ and $\sigma(\chi)$ are closely related; grey curves indicate the allowed regions of the joint distributions. The colored histograms show their observed distributions while solid curves indicate expectation: $\langle\chi^2\rangle=1$, $\mu(\chi)=0$, and $\sigma(\chi)=1$.
	}
	\label{fig:residual_stats}
\end{figure*}

A number of residual statistics are also included that provide valuable insight into the goodness-of-fit of a given model for a given source and band. In order to minimize contamination with neighbours, we consider only the pixels belonging to the source segment in the computing these estimates (same as in the decision tree). The primary statistic is $\chi^2$, already discussed in Section~\ref{sec:make_models}. Three other related statistics are produced by measuring the moments of the inverse variance weighted $\chi_i$ images where each $i$ pixel value indicates the significance of the residual in units of per-pixel uncertainty $\sigma_i$: the median $\mu(\chi)$, standard deviation $\sigma(\chi)$, and D'Agostino's $K^2$ test which measures the normality of the residual by combining estimates of skew and kurtosis\footnote{The $K^2$ test is generally stable only for sources which have more than 8 pixels in their segment.} \citep{dagostino_1970, dagostino_1990}. These statistics can also be combined to separate reliable models from poor fits and blends, as shown in Figure~\ref{fig:residual_stats}. 

Once forced photometry is completed, \farmer{} appends the measurements to (a copy of) the existing model catalog. This can be done on a band by band basis, or for all bands simultaneously. Output includes fluxes, as well as other parameters including band-specific positions and shapes if the user has allowed them to vary. Residual statistics are also included for every source in each measurement band.

\farmer{} has an additional diagnostic ability to measure photometry of these known, detected sources with concentric circular apertures of various diameters specified by the user. This is especially useful for constructing profiles of the radial flux growth. Aperture photometry can be measured on the science images (to get basic comparisons with the profile-fitting measurements), and it can go further by forcing the same apertures on the residual image and weight images. Most interestingly, these apertures can be forced on model images constructed by realizing the entire group of models into pixel-space. The aperture fluxes can then be readily compared with fluxes measured on the same apertures on the science image. Similarly, apertures can be forced on single sources realized into the pixel-space of the image in complete isolation; measurements in large apertures be compared with the total flux reported by \tractor{}\footnote{However, if the model is severely truncated by being realized into an image whose dimensions are much smaller than the full extent of the model then the integrated flux in large apertures will underestimate the total, correct flux measured by the normalization coefficient.}. Together, these aperture measurements can help diagnose model inaccuracies and bias, providing an effective means to internally validate the results of \farmer{}. 

Diagnostic images can be incredibly useful. \farmer{} can be configured to produce pixel-level background and RMS maps in addition to source and group segmentation maps. Importantly, \farmer{} can realize the entire model library of a brick as a reconstructed pixel-level model image from which corresponding residual and weighted significance $\chi$ images can be produced. Since catastrophic failures can result in models spanning large regions of the reconstructed model images, \farmer{} allows the user to automatically filter models based on $\chi^2_N$ or axis ratio such that they are not included in the reconstructed model, residual, or $\chi$ images (especially useful for cleaning residuals when searching for undetected signal). Also, models with negative fluxes will create positive flux in residual images; these can also be automatically filtered. Although removing sources at this level introduces incompleteness, it is likely that the measurements of these problematic sources are not scientifically useful anyways. To account for the missing area, \farmer{} also provides an effective mask image which flags pixels belonging to removed sources according to their segment ownership and computes the effective area of that mask. Although laborious, this is an optimal system for precisely determining the effective area from which a cleaned sample has been selected. Caveats regarding these reconstructed images are discussed in Section~\ref{sec:assumptions}, below.

\section{Benchmark and validation}
\label{sec:validation}

In this section we test and validate the performance of \farmer{} using a set of simulated deep images with COSMOS-like properties.

\subsection{Construction of Mock Images}
\label{sec:make_mocks}

The construction of the mock images used here follows the approach presented in L. Zalesky et al. (in prep.). Images are created to include a number of realistic features. The noise in each filter is matched to the RMS measured on real images used in \citet{Weaver2022_catalog}. Galaxy-type sources are included with random positions and orientations using the open-source code \texttt{GalSim} \citep{Rowe2015_GalSim} via the \texttt{RealGalaxy} class, which allows the user to inject images of real sources observed by \textit{HST} in the COSMOS field. Unfortunately, the morphology of these sources is only available at the resolution of \textit{HST} in one filter (F814W). In order to simulate wavelength-dependent profiles, we use parametric model representations of these galaxies (bulge+disk composites), and give red spectral energy distributions to bulge components and blue spectral energy distributions to disk components; this is handled internally within \texttt{GalSim} by the 
\texttt{RealGalaxy} class. To ensure a realistic colors for each galaxy, we have cross-matched the \textit{HST} catalog internal to \texttt{GalSim} to the COSMOS2020 catalog, and re-scaled the flux in each band that we simulate to that of the matched source. 

The shape of the galaxy number counts is fixed by the internal \texttt{GalSim} catalog, and all we modify is the normalization, such that resolved galaxies comprise $\sim$ 2/3 of all sources at intermediate magnitudes ($20 < m_{\rm i} < 24.5$). The \texttt{GalSim} counts are incomplete beyond $m_{\rm i} \gtrsim 25$, and so we inject PSF-models with \tractor{}, assuming a constant PSF in each band. This is reasonable since \citealt{Weaver2022_catalog} showed that objects fainter than 24.5\,AB in COSMOS2020 are generally unresolved. This means that the injected point sources are a fair test of \farmer{}'s ability to identify resolved and unresolved sources: had we somehow injected realistically sized galaxy models they would appear as unresolved sources anyway and so \farmer{} would have rightly fit them as such. The fluxes of these point sources are tuned such that together with the galaxy sources, the total sample yields a complete sample in the HSC-$i$ band to 28.5\,magnitude. Colors of point sources are assigned by randomly selecting sources of similar flux (within $\pm$ 0.1 mag) from the COSMOS2020 catalog and scaling the flux in a given filter to match the color. Finally, the number counts are calibrated and scaled according to the number counts of the COSMOS2020 catalog and to those in the empirical mock catalog of \cite{Girelli2020_mock}.

Our optical and NIR images are simulated at the same scale as the images used in \citet{Weaver2022_catalog} (0.15$^{\prime \prime}$/pixel). Likewise, we also simulate the mid-IR \textit{Spitzer}/IRAC images at their native resolution of 0.6$^{\prime \prime}$/pixel and then use \texttt{SWarp} \citep{Bertin2010_swarp} to resample them to 0.15$^{\prime \prime}$. This step introduces correlated noise that affects the effective degrees of freedom of the model fits (see Section~\ref{sec:performance}).

Although the galaxies in our simulated images are parametric representations, it should be noted that real galaxies feature structures such as spiral arms and star-bursting regions that are not captured by these models. As such, the performance of \farmer{} for the brightest sources assessed on this simulation is likely overestimated compared to real galaxy images.

\subsection{Procedure}
\label{sec:procedure}

We follow the general procedure outlined in Section~\ref{sec:farmer}. For simplicity and to ease the interpretation of our tests, we adopt the input PSFs used to produce the simulated images. No backgrounds are subtracted. These two aspects of our procedure are functionally equivalent to perfect knowledge of the PSF and of the image background. Sources are detected on a $izK_s$ \texttt{CHI-MEAN} image created using \texttt{SWarp} \citep{Bertin2010_swarp}. We do not model on the detection image due to the combined, chromatically dependent PSF. For this reason, and to preserve our selection function, we determine our best-fit model types, positions, and shapes by jointing fitting the same bands that constitute our detection image ($i$, $z$, and $K_s$) using their respective PSFs. Models are assigned according to the same decision tree structure as described in Figure~\ref{fig:farmer_decision}. Sub-optimal model assignments such as assigning resolved models to many poin
t-like sources tends to produce a photometric bias which manifests as a plateau or sharp rise in source magnitude distributions (i.e. number counts) at the threshold in magnitude beyond which sources are generally unresolved. Therefore we tune the decision tree to produce smoothly increasing number counts, and then tune further by spot checking residuals of individual sources; determining in $\chi^2_N$ a Pointsource penalty of 0.1 and a resolved model similarity threshold of 0.2. The modelling stage is run, which assigns models and optimizes their parameters on a group basis. 

\begin{figure*}[t]
	\centering
	\includegraphics[width=1\hsize]{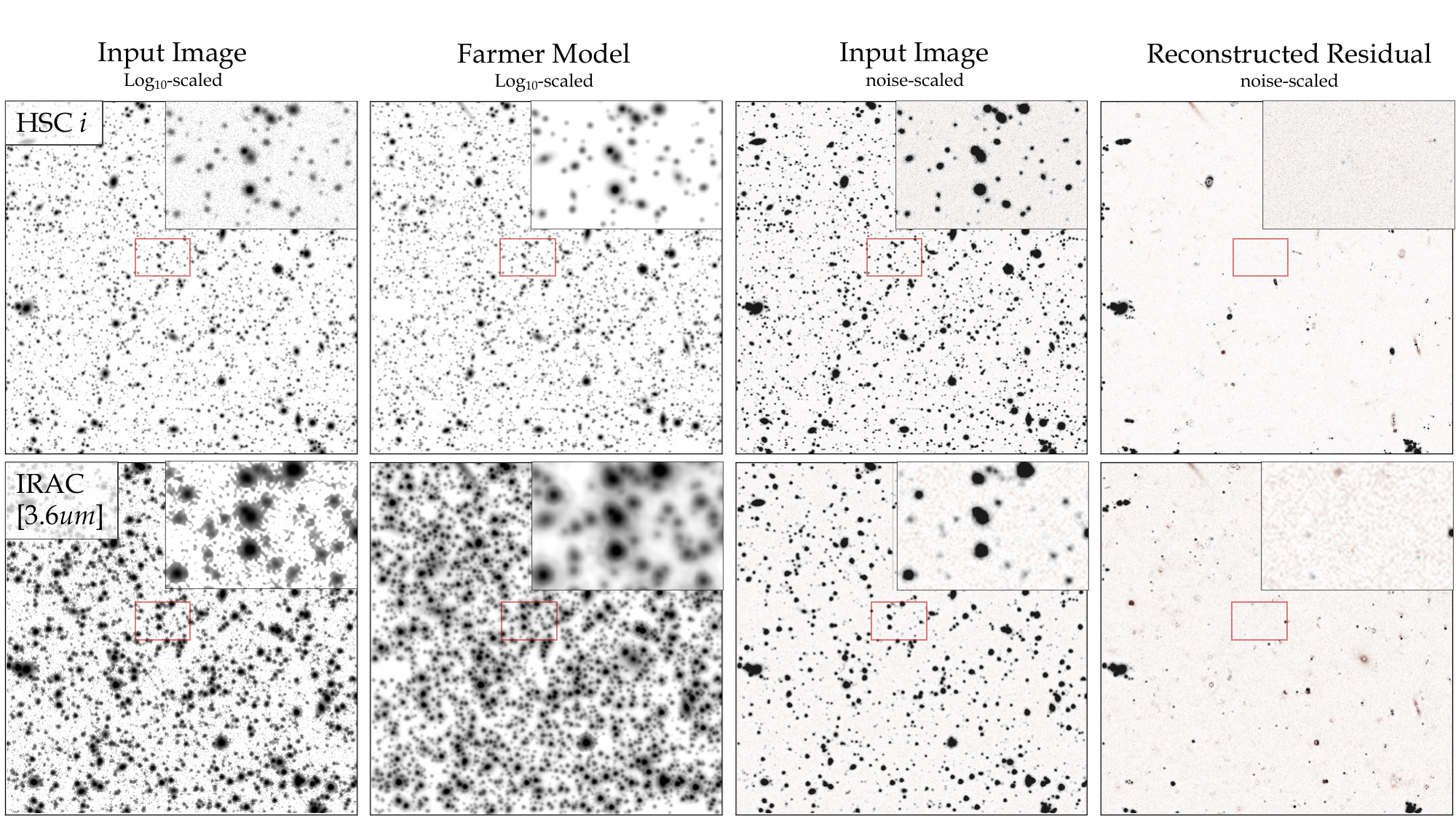}
	\caption{Results of forced photometry by \farmer{} on simulated fields of depths similar to COSMOS. Sources are $izK_s$-detected and modelled on $i$, $z$, and $K_s$ jointly, then forced on other bands including channel~1. Models can be compared to input images in the two leftmost panels in $\mathrm{log}_{10}$ scaling where morphology is visible. Insets show a zoom-in of a smaller region. Residuals can be compared to input images in the two rightmost panels in $\pm3\sigma$ scaling to highlight faint signal above (black) and below (red) the background. While some sources were skipped as they were too complex, other notable over-subtractions are driven by blends and/or injected sources that are not well-described by the five model types.}
	\label{fig:farmer_residuals}
\end{figure*}

We perform force photometry by re-fitting the models on the bands of interest: $r, i, z, K_s$, channel~1, and channel~2. Positions and shapes are fixed for each object, with only the five independent fluxes free to vary. Figure~\ref{fig:farmer_residuals} shows the reconstructed model images and residuals produced by \farmer{} over a region of the simulated $i$ and channel~1 mosaics. The vast majority of sources are well modelled with only a handful of failed fits which are left in the residual map. While the value of visual inspection of residuals cannot be understated, a rigorous statistical analysis can provide powerful quantitative insight.

\subsection{Model and Decision Tree Performance}
\label{sec:performance}

Now we use the suite of statistics provided by \farmer{} to assess the performance of the models and decision tree.

\begin{figure}[t]
	\centering
	\includegraphics[width=1\hsize]{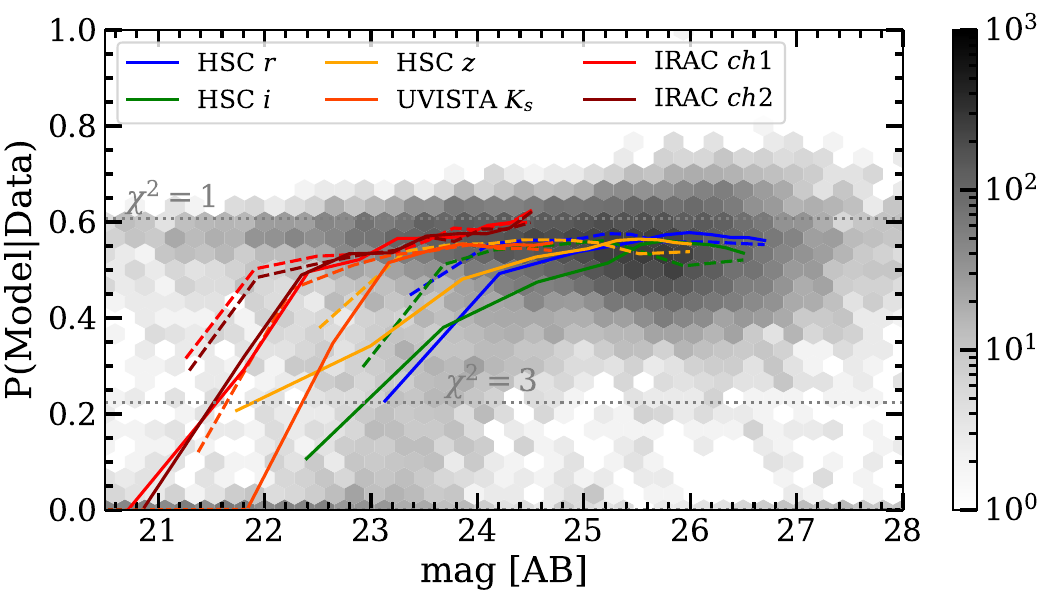}
	\caption{Probability of models as a function of apparent magnitude. Results from $i$ are shown in grey histograms, summarized by binned medians of unresolved (solid) and resolved models (dashed). Other band are similarly summarized by binned medians.
	}
	\label{fig:model_probability}
\end{figure}

As demonstrated by Figure~\ref{fig:model_probability}, the probability of the model given the data (inversely proportional to $\chi^2_N$) is greatest for faint sources across all bands. For images of high spatial resolution (e.g. $r$, $i$, $z$), the model performance degrades for both resolved and unresolved models at magnitudes brighter than $\sim 24\,\mathrm{AB}$, although with considerable variance. These bright sources are smooth in our simulations, however they are still more complex than the models supplied by \tractor{}. Additionally, brighter sources usually subtend a larger area and so reside in more complex groups where blending makes accurate photometry more challenging. A notable exception are bright point-like sources which are typically well-fit by the PointSource model type.

The NIR and IR bands ($K_s$ and IRAC) have slightly better performance at bright magnitudes. This is because their resolution threshold is at a brighter magnitude and so these particular bands contain a higher fraction of bright sources which appear unresolved. Whether or not \farmer{} assigned resolved or unresolved models to these sources, the resolution is low enough that they are effectively unresolved. Photometry is then made easier because there is little dependence on accurate model shapes. The key insight therefore is that the effectiveness of profile-fitting photometery is not dependent on source magnitude directly, but rather on the size of the source and whether or not is is resolved, with some lesser dependence on the resolution of the bands used to derive the models.

\begin{figure}[t]
	\centering
	\includegraphics[width=1\hsize]{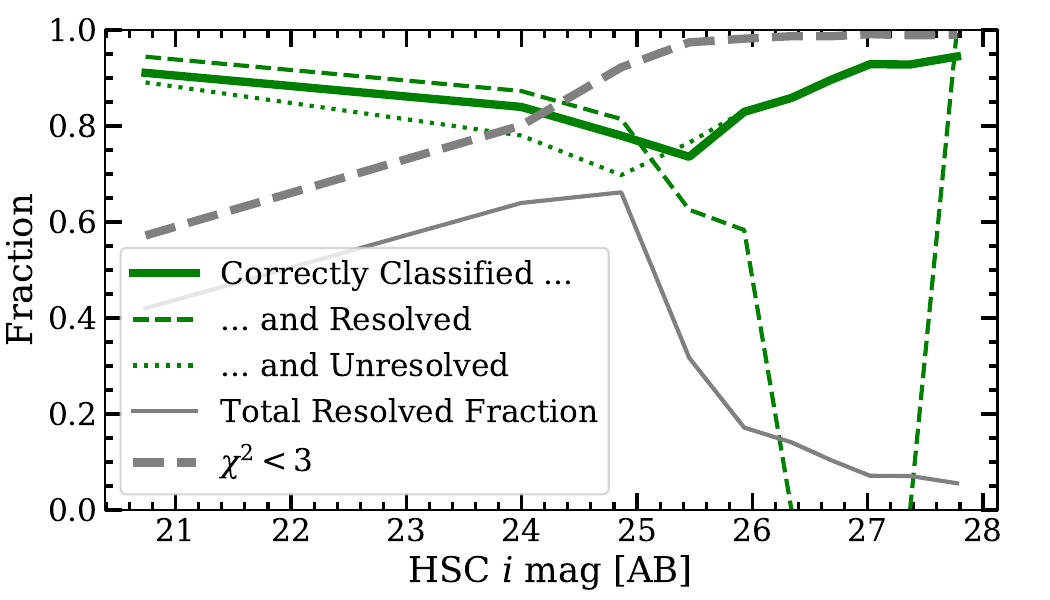}
	\caption{Fraction of unresolved (resolved) sources correctly assigned an unresolved (resolved) model by \farmer{} as a function of apparent $i$-band magnitude is shown by the solid green curve, and broken down into resolved and unresolved subsets by the dashed and dotted curves, respectively. The in recovery fraction at $i\sim25$ is expected as sources at that brightness are typically only marginally resolved and so challenge the decision tree. We consider only sources with $\chi^2<3$ in the $i$-band as they are considered reliable, the fraction of which is shown by the grey dashed curve.  Our simulated field  at $i>25$ uses mostly point-like sources to reflect real conditions in a COSMOS-like survey; the total resolved fraction is shown by the solid grey curve.
	}
	\label{fig:recovery_fraction}
\end{figure}

As shown in Figure~\ref{fig:recovery_fraction}, \farmer{} is generally able to correctly assign resolved models to sources which are injected as resolved galaxies, and unresolved models to those which are injected as unresolved point sources (which could be stars or galaxies -- \farmer{} does not try to separate them). As alluded to earlier, the resolution threshold averaged over the modelling bands ($\sim25$ for $izK_s$) is where it is most difficult for \farmer{} to distinguish between resolved and unresolved sources and so ultimately the fine tuning of the decision tree is aimed at improving performance in this regime. Based on our tuning, \farmer{} correctly assigns $>75\%$ of marginally resolved sources. 

While it appears that \farmer{} is not able to correctly assign resolved models to injected resolved galaxies at $i>25$, this is almost certainly because these sources actually appear unresolved in our simulated images. It should be noted, therefore, that while a given source in the simulated images corresponds to either a resolved galaxy or unresolved point source model, the former may be be effectively unresolved in the image if it is smaller than the PSF. Identifying such cases in the $i$, $z$, and $K_s$ bands is therefore of interest as \farmer{} should not be expected to assign them a resolved model. These cases cannot be cleanly identified beforehand, nor is it possible to identify them afterwards with full confidence. As a result, the performance of \farmer{} may be expected to be better than it appears in Figure~\ref{fig:recovery_fraction} around the $i\sim25$ resolution threshold.

\begin{figure}[t]
	\centering
	\includegraphics[width=1\hsize]{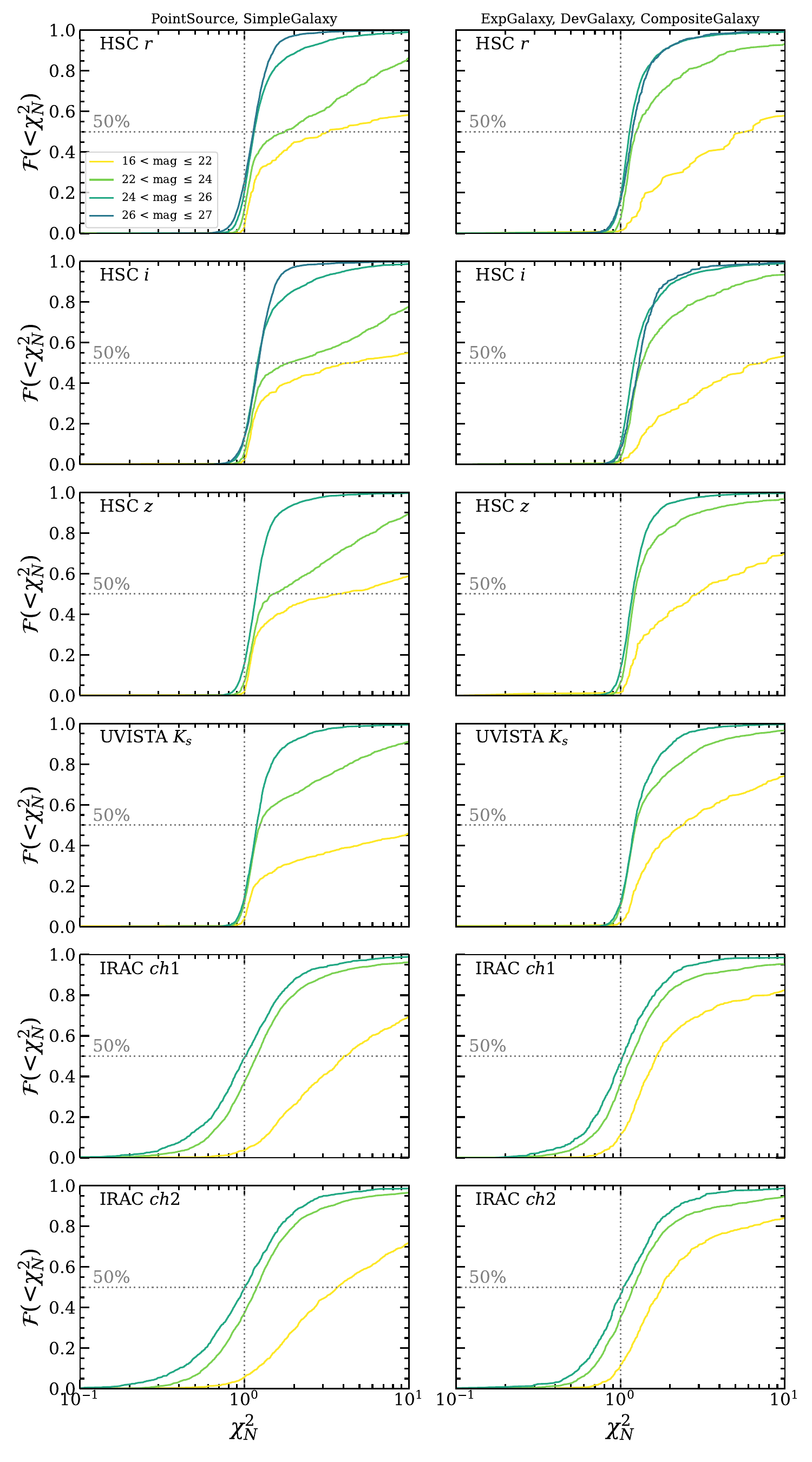}
	\caption{Fraction of sources below a certain $\chi^2$ as a function of band and magnitude for unresolved (left) and resolved models (right).
	}
	\label{fig:cdf_chi2}
\end{figure}

The performance of models optimized in forced photometry is also generally better at faint magnitudes where sources are typically unresolved. Figure~\ref{fig:cdf_chi2} shows the fraction of sources below a given reduced $\chi^2_N$ in four ranges of magnitude for each band separated into unresolved and resolved model types. A sample which is $\chi^2$ distributed reduced by $N$ degrees of freedom should have an expectation value of unity. Its cumulative distribution should therefore be approximately evenly divided around $\chi^2_N\approx1$. It should be noted that $\chi^2$ is a measurement of significance and is therefore dependent on accurate per-pixel errors.

The performance of models for the well-resolved bands ($r$, $i$, $z$, $K_s$) is better for faint sources irrespective of resolved or unresolved models. Overall these distributions seem slightly shifted towards larger values of $\chi^2_N$.  Inspection of the residuals suggest these models are well fit, and so this shift may be due to inaccurate per-pixel errors, or pixel covariance which is not accounted for by $\chi^2$ which assumes independent, Gaussian distributed data. For bright sources, a tail develops at $\chi^2_N>10$ which also suggests an increased fraction of bad models. This is expected as any imperfection in the model will add some term proportional to the square of the source flux. By inspection, we confirm that the complexity of the injected galaxies is not always well-captured by the smooth models from \tractor{} (as would happen in real images). Source crowding may also play a role for these typically large, bright sources that may have fainter sources near their wings that if not detected may cause a photometric bias. 

The two infrared bands (channel~1 and 2) appear to have slightly better performance at faint magnitudes. There does not seem to be a shift, which relative to the bluer bands may be due to greater degree of signal covariance relative to the bluer bands (from the larger PSF) whereby a good fit in one pixel means one can expect to achieve a good fit in the adjacent pixels. A tail does not develop for bright sources, which instead are shifted towards higher $\chi^2$. This systematic behavior suggests that \farmer{} has the greatest difficulty modelling the bright IRAC sources in general. This is not a surprise given that the IRAC images have worse resolution, meaning that light from neighboring (bright) objects can impact sources in a given group. Because this extra light is not expected by the group model, it may lead to a photometric bias.

\begin{figure*}[t]
	\centering
	\includegraphics[width=1\hsize]{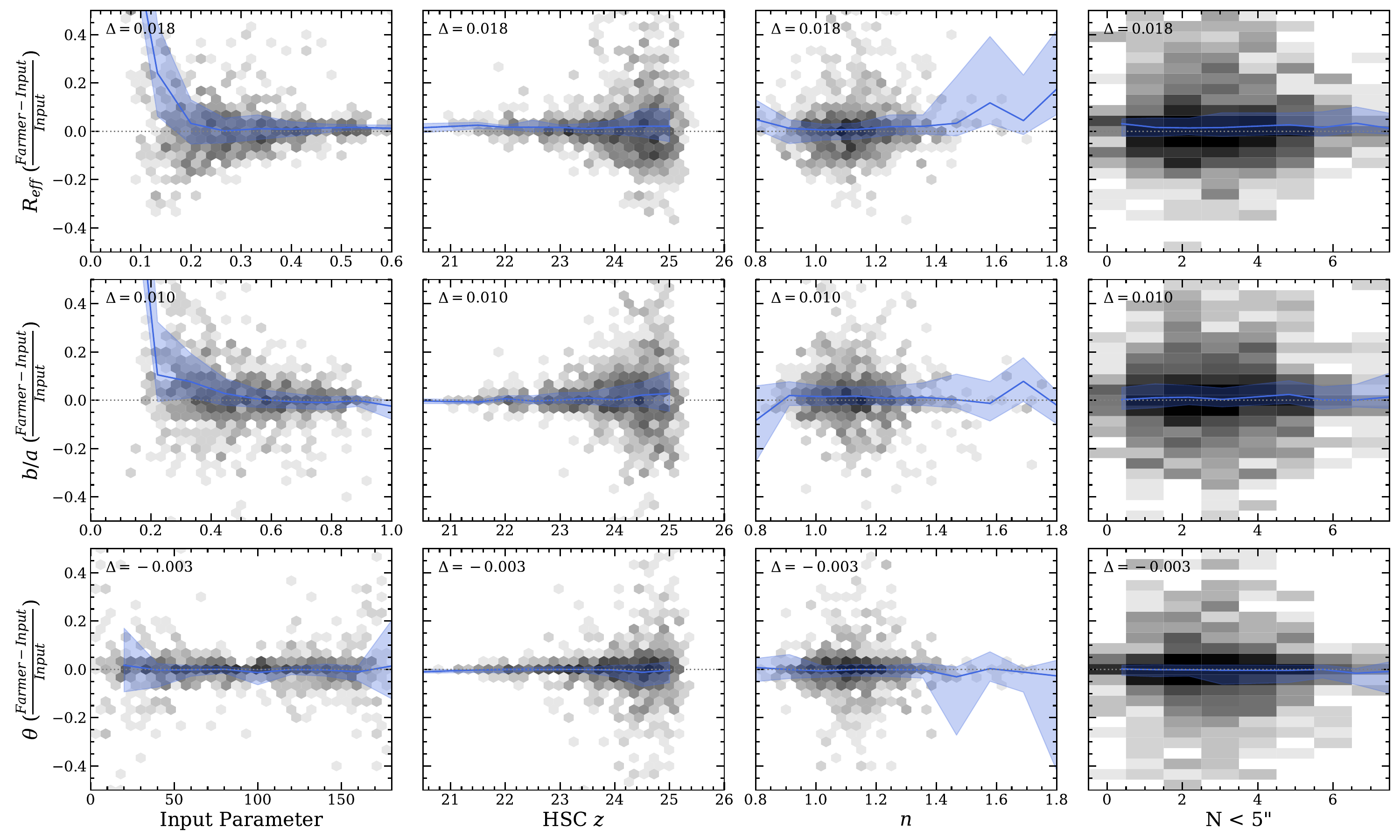}
	\caption{Recovery of effective radius ($R_{\rm eff}$; top row), axis ratio ($b/a$; middle row) and position angle ($\theta$; bottom row) as a function of input parameter, $z$-band magnitude, S\'{e}rsic index ($n$), and the local source density (number of sources within 5\arcsec). Greyscale distributions are summarized by binned medians (colored curves) with 68\% range indicated by the envelope.
	}
	\label{fig:shape_recovery}
\end{figure*}

\farmer{} also provides accurate shape measurements for all resolved sources. Figure~\ref{fig:shape_recovery} demonstrates the recovery of axis ratio and position angle of the simulated galaxies, finding agreement within 1 per cent. There are no obvious biases in any parameter, whether compared to itself, source magnitude, S\'{e}rsic index, or local source density. The only notable deviations are expected: circular sources with $b/a\sim0$ where the axis ratio signal is very weak and small sources where $R_{\rm eff}$ approaches the pixel scale of the image (0.15\arcsec/px). The insensitivity to local source density gives \farmer{} a considerable advantage over shapes estimated from \texttt{Source Extractor}.

\subsection{Counts and photometric accuracy}
\label{sec:counts}

Credible survey science ultimately rests on a foundation of complete samples and accurate photometry. We characterize the relevant performance of \farmer{} in the following assessments.

\begin{figure}[t]
	\centering
	\includegraphics[width=1\hsize]{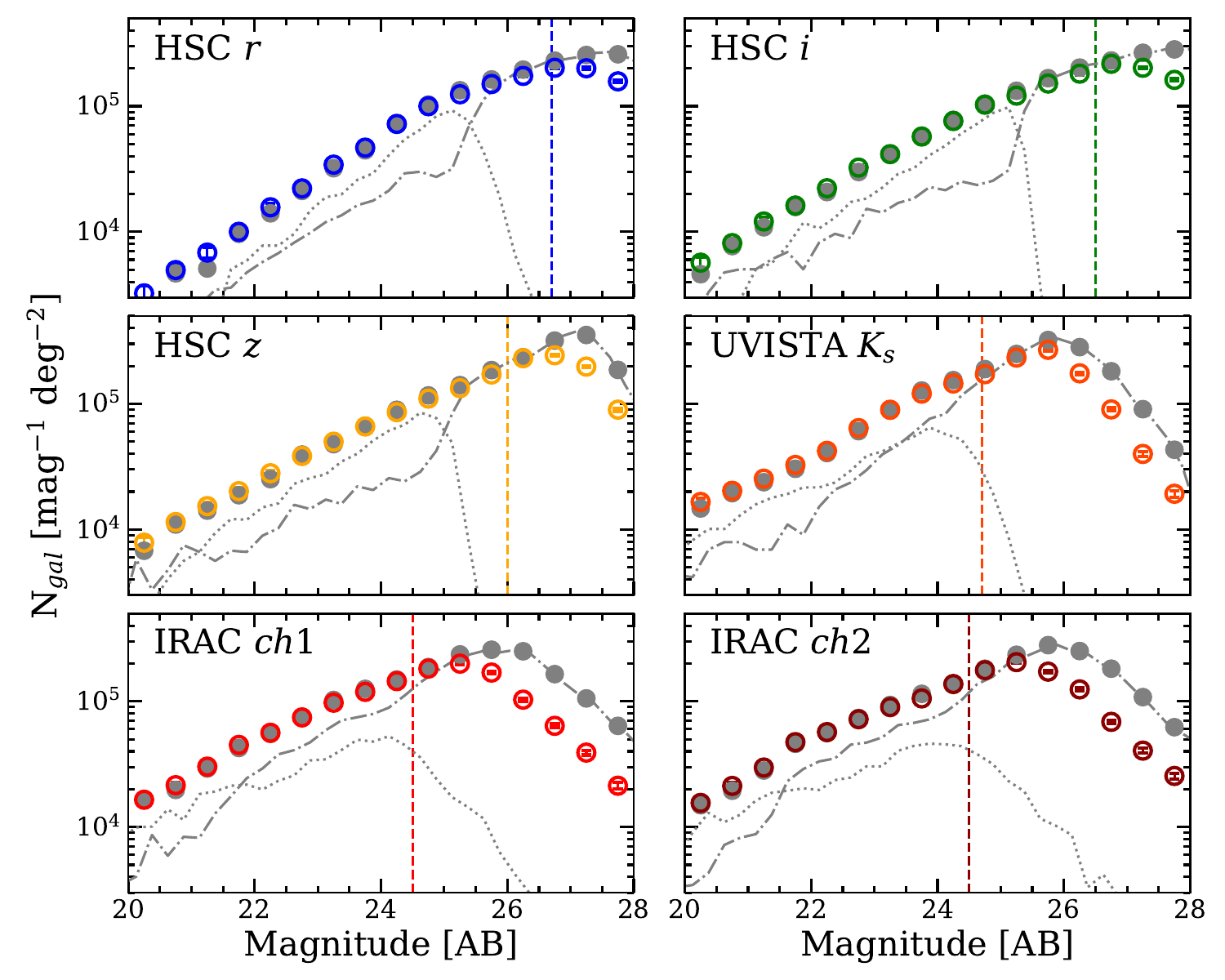}
	\caption{Number counts are shown for each band corresponding to the simulation input for all sources (filled grey points), resolved sources (grey dotted curve), and unresolved sources (grey dash-dot curve). This is compared to output from \farmer{} (unfilled colored points with Poisson uncertainties) for an $iz$-selected sample. Nominal depths are shown by the vertical colored lines.
	}
	\label{fig:number_counts}
\end{figure}

Source number counts not only diagnose issues in sample selection and incompleteness, but are also sensitive to photometric accuracy. The number counts of injected sources in our simulated images are shown alongside those recovered by \farmer{} in Figure~\ref{fig:number_counts}. The recovery of number counts is generally excellent. They are complete up the limiting magnitude of each band, which is most important for the $i$, $z$, $K_s$ bands used in sample selection as incompleteness in other bands may be driven by selection effects. For instance, a small fraction of faint $r$-band sources is missing from our sample as expected given the simulation includes real galaxy colors and these predominantly blue sources are likely faint in our redder detection image. We can trust that \farmer{}'s decision tree is performing well given that there are no extended plateaus or sharp rises present anywhere in the number counts, in combination with other available diagnostics (e.g., residuals, $\chi^2$, etc.).

\begin{figure}[t]
	\centering
	\includegraphics[width=0.92\hsize]{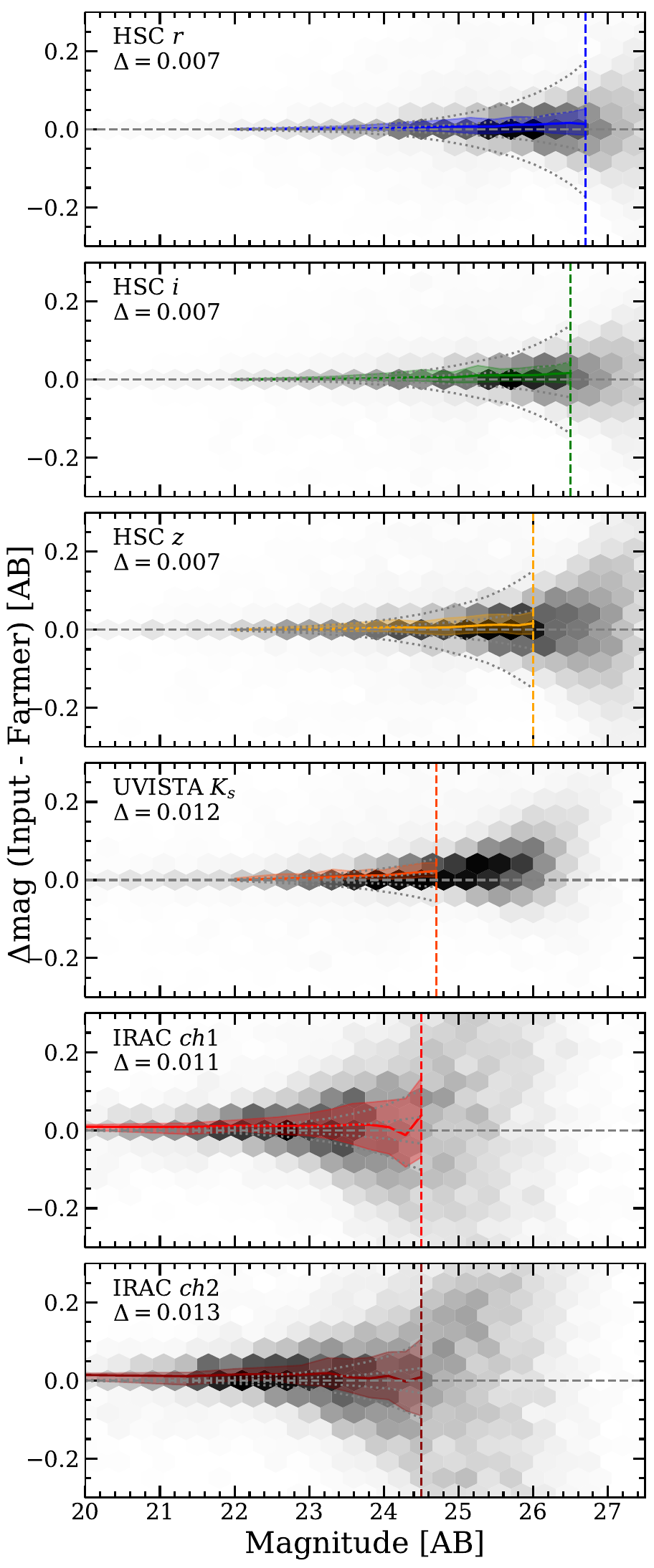}
	\caption{Photometry produced by \farmer{} is compared with true fluxes of simulated sources for all bands. Differences in magnitude as a function of input magnitude (grey histograms) are summarized by binned medians (colored curves) with 68\% ranges indicated by the colored envelope out to the nominal depth limit of each band (vertical colored lines). Expected $\pm1$ and $3\sigma$ uncertainties on $\Delta\mathrm{mag}$ are computed as medians from the \farmer{} uncertainties (grey dotted curves).
	}
	\label{fig:deltamag}
\end{figure}

The most important measurement is ultimately photometry. As shown in Figure~\ref{fig:deltamag}, the photometry measured by \farmer{} is seen on median expectation to be accurate below 0.05\,AB in all bands, including IRAC. There are no significant systematic biases, with only a small trending towards overestimated fluxes for faint sources in $K_s$. The 68\% scatter is similar to the typical magnitude uncertainty at a given magnitude for $r$, $i$, $z$, and $K_s$. For IRAC bands, the scatter is about three times larger than the typical magnitude uncertainty, suggesting that the photometric uncertainties may be underestimated. This may be expected given the high spatial covariance of noise in IRAC images. 

\begin{figure}[t]
	\centering
	\includegraphics[width=1\hsize]{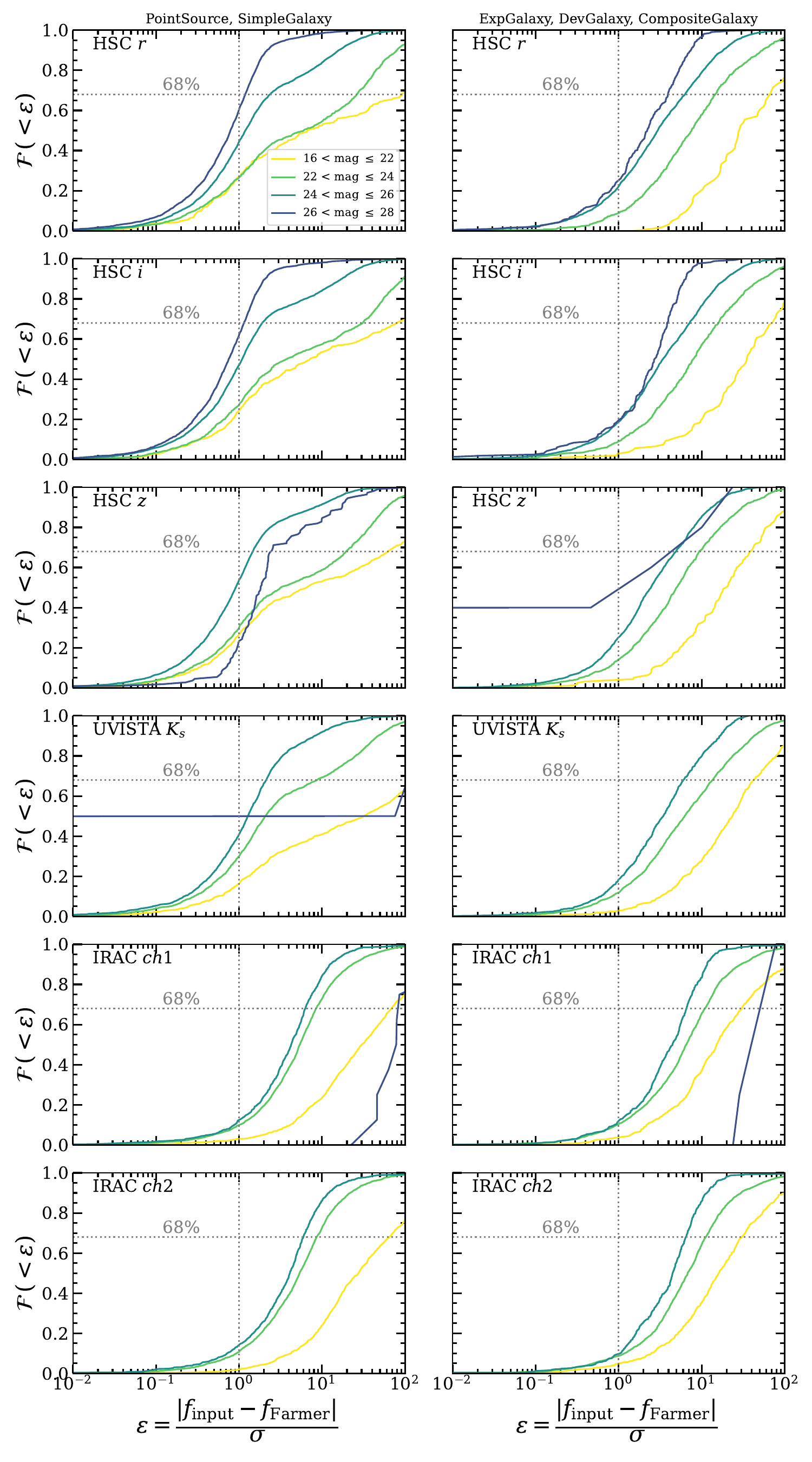}
	\caption{Fraction of sources whose relative photometric error $\epsilon$ is less than a certain value, broken down by resolved (left panels) and unresolved models (right panels) for each band. On expectation, $|\epsilon<1|$ for 68\% of sources where a significant departure may indicate under- or over-estimation of photometric uncertainties.  
	}
	\label{fig:cdf_ztest}
\end{figure}

Photometric measurements are more appropriately assessed by directly examining the cumulative distributions (CDF) of relative error $\epsilon=|f_{\rm input} - f_{\rm Farmer}|/\sigma$. These are shown in Figure~\ref{fig:cdf_ztest} broken down by band and separated into resolved and unresolved model types. On expectation, 68\% of sources should be contained by $|\epsilon|\leq1$. Given the lack of bias in our photometry, deviations of the $\epsilon$ CDFs from this expectation can be directly attributed to innapropriate flux uncertainties resulting from miscalibrated weights and/or spatially covariant noise. 

We see a similar picture to the $\chi^2$ CDFs in Figure~\ref{fig:cdf_chi2} whereby photometry of faint sources measured in the high spatially-resolved bands ($r$, $i$, $z$, and $K_s$) better follows expectation compared to photometry of bright sources. The distribution of $\epsilon$ for bright point sources has a tail as even the smallest biases are expected to yield large $\epsilon$ values as the typical flux uncertainties are small. However, the same is not true for the resolved models which are systematically shifted towards larger $\epsilon$ with increasing brightness. This may suggest poor modelling performance of the brightest sources, in accord with previous results.  

The $\epsilon$ CDFs for the IRAC bands are significantly shifted towards higher values in agreement with the results from Figure~\ref{fig:deltamag}. This is further evidence that the weights from our IRAC mocks may produce underestimated photometric uncertainties. This is not an immediate confirmation, however, because both $\chi^2$ and $\epsilon$ assume independent, Gaussian distributed data which may not be the case in instances of significant pixel covariance; e.g. as in the case of IRAC as it has been up-sampled such that the PSF is correlated across more pixels. While this is treated to some degree by \texttt{SWarp}, the resulting weights seem to still be overestimated.

\begin{figure}[t]
	\centering
	\includegraphics[width=1\hsize]{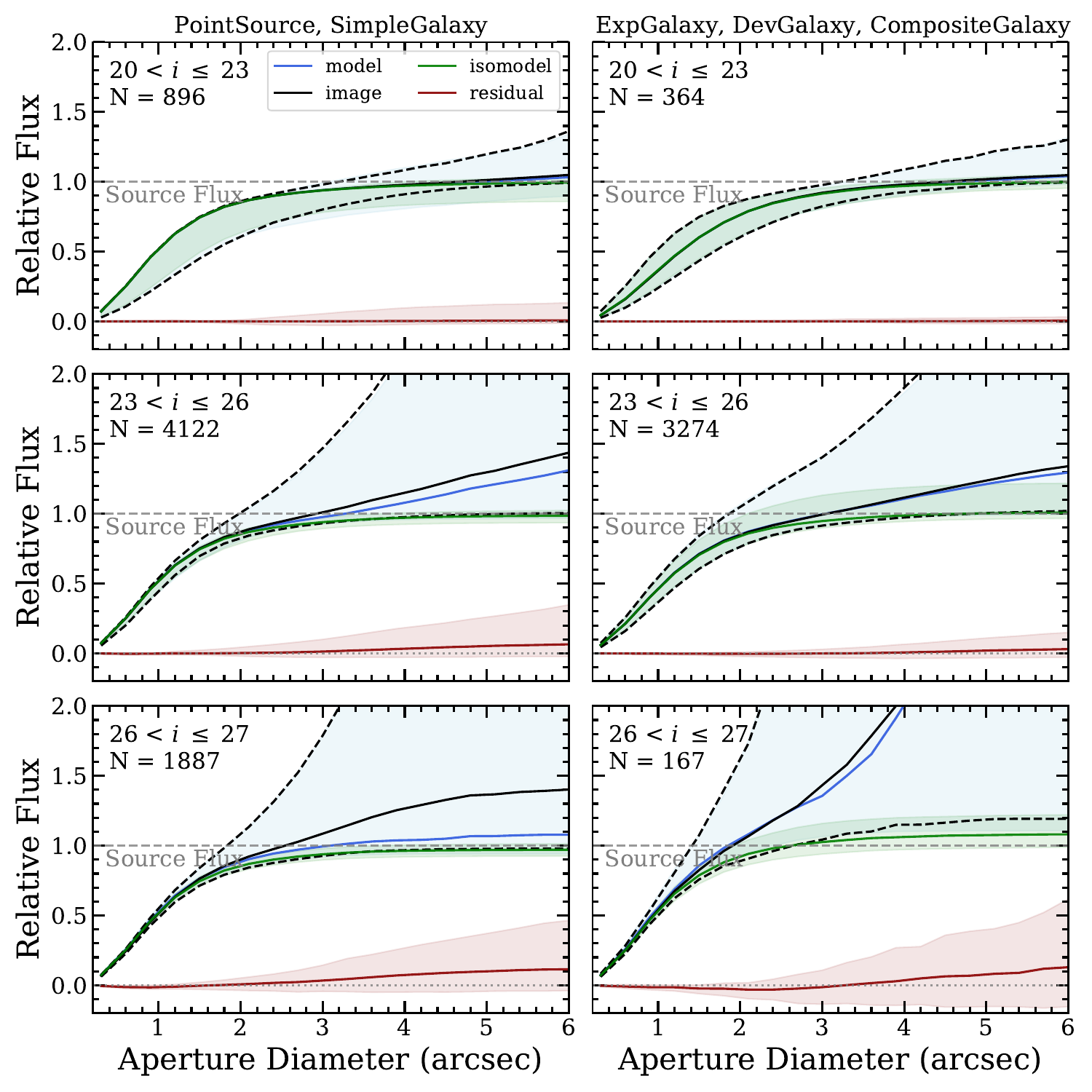}
	\caption{The accuracy of the models in various magnitude regimes can be assessed by measuring fluxes are in circular apertures of increasing diameter on simulated input $i$-band image (black), \farmer{} joint model with neighbors (blue), their residual (red), and \farmer{} model without neighbors (green). For easy comparison, the y-axis shows each aperture flux measurement normalized to the true input flux and are summarized by a median and 68\% range, which are shown as colored regions except for the input image which is shown as black dashed lines. While faint sources contain higher fractions of blends at a given radius (the black rises above unity in the lower panels), \farmer{} is still able to recover the input source flux (green tends to unity at large radii).
	}
	\label{fig:aper_diagnostic}
\end{figure}

Another way to investigate typical model accuracy is demonstrated in Figure~\ref{fig:aper_diagnostic}. As described in Section~\ref{sec:catalogs}, \farmer{} can be configured to extract flux in circular concentric apertures at every source position. We have measured fluxes in several aperture sizes with sub-arcsecond steps for both resolved and unresolved models computed on the group images, models, and residuals. Fluxes are also measured consistently for each source individually, such that they are realized in isolation of other sources (the `isomodel'). The largest aperture is 6\arcsec{} in diameter which likely captures flux from neighboring sources in the $i$-band image used here. As expected, while the `image' and `model' flux grow beyond the input source flux due to the presence of neighbors, that of the `isomodel' tends towards agreement with the input source flux (i.e. 1), and that of the residuals generally tends towards zero.

Bright sources are typically large on the sky such that the largest apertures are dominated by the bright source with insignificant contributions from faint neighbors. The apertures measured on the image, model, and isomodel agree well for both resolved and unresolved bright sources, and tend towards agreement with the true input flux at large radii (a value of 1 on the y-axis). Interestingly, the flux at small radii varies significantly. This is driven by the variation in light profiles (i.e. S\'ersic index) that's more visible for bright, well-resolved sources. In the case of sources fit with PointSource or SimpleGalaxy models, the variation is driven entirely by the different light profiles. As one might expect, including only sources fit with PointSource models results in almost no variation whatsoever as all point sources have the same curve of growth.

The behavior is different for fainter sources. While their image and model fluxes continue growing even at large apertures, the flux of the isomodel stops growing around 3\arcsec{} as no new flux is captured by the apertures and agrees with the true input flux. The situation changes again for the faintest sources where on average there is blending at radii smaller than 3\arcsec{} as shown by the divergence of the black image and blue model flux growth curves from that of the isomodel in green that on average agrees with the true input flux. Hence, while there is blending of sources within even 2-3\arcsec apertures in $i$-band, the approach used by \farmer{} produces fluxes which are not typically affected by blending\footnote{This will not be true in cases where blended sources are not separated by detection, see Section~\ref{sec:assumptions}.}.

\subsection{Deblending in IRAC}

\begin{figure}[t]
	\centering
	\includegraphics[width=1\hsize]{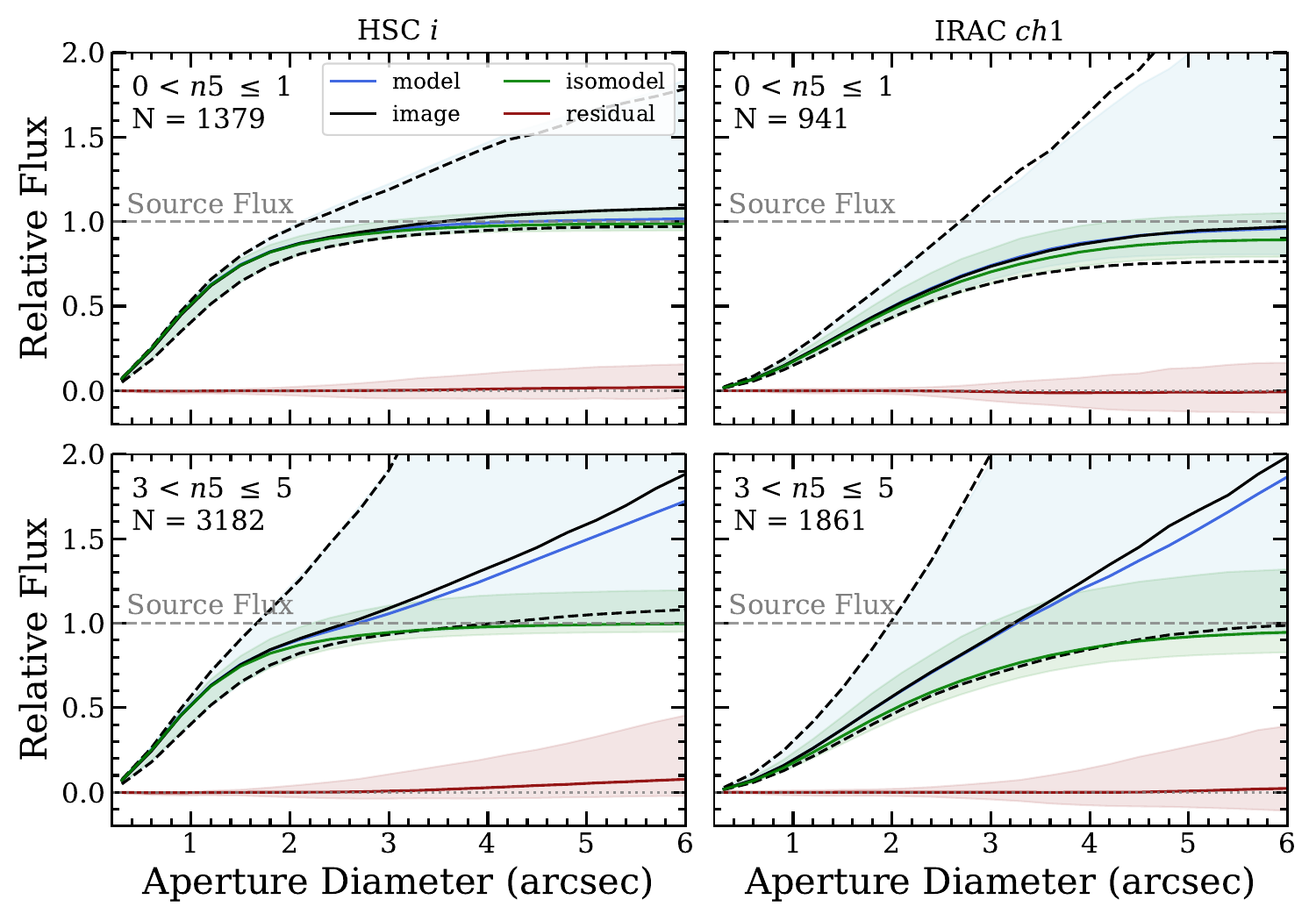}
	\caption{Summary of de-blending power of \farmer{}. Similar layout to Figure~\ref{fig:aper_diagnostic}, but for sources photometered in $i$-band (left) and channel~1 (right) broken down by local density $n5$ defined by the number of sources within 5\arcsec.  
	}
	\label{fig:aper_density}
\end{figure}

Here we assess the de-blending performance of \farmer{} more thoroughly in the context of our simulated IRAC images in Figure~\ref{fig:aper_density}.

Similar to Figure~\ref{fig:aper_diagnostic}, photometry is measured in apertures forced on source positions computed on the images, models, isomodels, and residuals. As a baseline, growth of flux for sources measured in $i$ are in agreement between the images, model, and isomodel, as well as the true total flux for isolated sources. However, for sources in crowded regions the flux measured on the image and model continues to grow whereas that of the isomodel flattens out around 4\arcsec\,in agreement with the true flux. 

Although IRAC images have very different properties compared to HSC's $i$ band, the behavior for isolated sources is similar. The only difference being that larger apertures are required to encompass the total flux of IRAC sources. Aperture photometry measured in crowded regions of IRAC images, however, quickly become contaminated by the flux of neighbors so that no aperture diameter can cleanly measure the total flux of the central source. While the encompassed flux from both the image and model apertures grows exponentially, that of the isomodel finds good agreement with the true total flux of the simulated source. What is incredible is that the flux growth curve of the isomodels in green deviates from that of the total group images in black and their joint models in blue already below 2\arcsec, meaning that de-blending is typically significant in our IRAC images even on these small scales. As such, the only tenable way to obtain accurate, high signal-to-noise photometry of IRAC sources is with a profile-fitting approach which, crucially, provides for joint modelling with neighboring sources as employed by \farmer{}.

\section{Summary and outlook}
\label{sec:summary}

While deep galaxy surveys from space-based facilities offer exquisitely resolved images, ground-based surveys are capable of efficiently obtaining similar depths over significantly larger areas where searches for rare populations can be conducted, although at the cost of resolution. Already such survey images contain source densities that demand increasingly smaller aperture photometry to avoid crowding, which results in more uncertain measurements \citep{laigle_cosmos2015_2016, Weaver2022_catalog}. As we have demonstrated, aperture photomery will grow less reliable as extragalactic fields deepen and become more crowded. Investments in deep, ground-based surveys will continue in the coming decade and so it should be expected that the magnitude of these challenges will only increase. Profile-fitting methods have been a longstanding technique for measuring low-resolution infrared images as they are less susceptible to source crowding. However, their advantages are now needed in the optical and near-infrared regimes. \farmer{} attempts to answer this call.

We have explored the methodology of \tractor{} whose photometry does not require that images be PSF homogenized, and total fluxes are reported solely based on the scaling of the model profile; avoiding the need for often ill-posed aperture corrections. However, we highlighted several obstacles preventing us from directly applying \tractor{} to deep, crowded galaxy fields. These problems were solved by developing \farmer{} which leverages an efficient albeit complex decision tree to assign models to sources in an optimal and less pathological way compared to simpler approaches. The decision tree is shown to be more than a useful algorithm, but indeed a required development in overcoming challenges related to blending in deep fields. \farmer{} is also a means by which to organize survey data so that one can utilize massively parallelized computing facilities to streamline computational time from potentially years down to only a few weeks. Profile-fitting photometry is, however, more complicated than apertures and comes with its own drawbacks and considerations ranging from selection functions to image resolution, and from deblending capabilities to computational limits. 

In a series of validation tests, we examined the ability of \farmer{} to photometer sources in realistically simulated images. We found no significant biases in photometry in any band. Furthermore, we illustrated the unique advantage of \farmer{} in de-blending sources in low-resolution images like IRAC. Still, bright and potentially resolved sources will continue present a limitation when employing smooth model profiles. On the other extreme, \farmer{} has been shown to provide incredibly accurate photometery of the faintest unresolved sources, and in this sense it helps open the door to the distant Universe.

Still, challenges in profile-fitting photometry remain and many difficult problems are yet unsolved. While we have demonstrated that \farmer{} will provide accurate photometry for the next generation of deep, crowded fields, we must continue to innovate as we move towards deeper and more complex surveys promising even greater discoveries.

\farmer{} is available to download from \href{https://github.com/astroweaver/the_farmer}{GitHub} and Zenodo: \dataset[doi:10.5281/zenodo.8205817]{https://doi.org/10.5281/zenodo.8205817} \citep{john_weaver_2023_8205817}.




\begin{acknowledgements}
The authors thank 
Dustin Lang, Charles Steinhardt, Keith Horne, and Ranga-Ram Chary
for helpful discussions. We are also grateful for the many helpful and constructive comments from the anonymous referee. 

The Cosmic Dawn Center (DAWN) is funded by the Danish National Research Foundation under grant No. 140. ST, GB and JW acknowledge support from the European Research Council (ERC) Consolidator Grant funding scheme (project ConTExt, grant No. 648179). OI acknowledges the funding of the French Agence Nationale de la Recherche for the project ``SAGACE''. HJMcC acknowledges support from the PNCG. ID has received funding from the European Union’s Horizon 2020 research and innovation programme under the Marie Sk\l{}odowska-Curie grant agreement No. 896225.
This research is partially funded by the Joint Survey Processing effort at IPAC/Caltech through NASA grant NNN12AA01C. The HST COSMOS program was supported through NASA grant HST-GO-09822. More information on the COSMOS survey is available at \url{https://cosmos.astro.caltech.edu/}. This work used the CANDIDE computer system at the IAP supported by grants from the PNCG and the DIM-ACAV and maintained by S. Rouberol.

\software{\texttt{numpy} \citep{numpy2011},
\texttt{matplotlib} \citep{matplotlib2007},
\texttt{astropy} \citep{2013A&A...558A..33A, 2018AJ....156..123A},
\texttt{Source Extractor} \citep{bertin_sextractor:_1996},
\texttt{\texttt{PSFEx}} \citep{bertin_psfex_2013},
\texttt{SWarp} \citep{Bertin2010_swarp},
\texttt{GalSim} \citep{Rowe2015_GalSim}, and
\texttt{The Tractor} \citep{lang_tractor_2016}
}
\end{acknowledgements}


\bibliography{thefarmer.bib}


\appendix

\section{Considerations, assumptions, and limitations}
\label{sec:assumptions}

Although \farmer{} effectively extends the functionality of \tractor{} to include source detection and grouping, model assignments, catalog creation, and computational efficiency, these advantages come with considerable limitations that are discussed below.

\subsection{Image Preparation \& Source Grouping}
\label{sec:limit_detection}

Several aspects of the image preparation and group identification stages are unique to \farmer{}. 

\textit{How should one determine how many bricks should a mosaic be broken into? }
This is primarily a computational concern. We can understand why by considering the combined perimeter of all the bricks in a mosaic of a fixed size; the perimeter is large when bricks are small and vice versa. The larger the perimeter, the greater the chance that the brick will split across a group of sources which should be ideally modelled simultaneously in the same brick. In general this should be avoided, and so bricks should be made as large as possible. It is possible for each brick to be operated on by \farmer{} independently, which means they can be parallelized across computational nodes or even processed by different computing facilities altogether. The brick also needs to be read into memory at runtime, and so should be sized appropriately for the memory capacity of a given computational facility. Bricks from forced photometry are typically the largest files as they contain all bands of interest, their weight maps, and masks and so can become tens of gigabytes for even modest dimensions.

\textit{What about sources near the edges of bricks that extend into the next brick?}  It is up to the user to determine how large the brick overlap should be. In general, the overlap should be large enough that the largest sources of interest, placed at the brick edge, would not extend beyond the overlap. While one can set a large brick overlap, doing so comes at the cost of memory and computational overhead. Excessive brick overlaps should be avoided where possible. 


\textit{How should one assess if groups are correctly identified?} 
As discussed in Section~\ref{sec:detection}, groups of sources are identified by joining source segments that have been dilated by some morphological structure whose size dictates the extent of the dilation. The segments are constructed from the detection stage, and so one should only consider the detection image when assessing the identification of groups. Dilation is necessary because the segmentation extent in \texttt{Source Extractor} (and \texttt{SEP}) is tied to the significance level set for the detection. In some cases the segments may not capture the full extent of two neighboring sources such that they should be simultaneously modelled, but their segments do not touch. Hence the size of the dilation structure should be set so that these kinds of nearby sources are correctly assigned into one group. This is most easily assessed by inspection, and tuned in successive trials.

It is important to note that morphological dilation can destroy segments nearby larger ones. For this reason the dilation is carried out on a copy of the segmentation image which has been made binary such that pixels assigned to the background are set to zero, and those active pixels assigned to sources are set to 1. Segments which are already touching are now indistinguishable, and the dilation simply enlarges the footprint of contiguous regions of active pixels. The small segments remain identifiable from the segmentation image. This is important, because the group pixels belong to the group itself; no one group pixel belongs to a single source. That ownership is retained in the original segmentation only. This is essential because while the joint likelihood maximized by \tractor{} is computed over the group pixels, \farmer{} can still judge the fit of individual sources from the $\chi^2$ computed over their uniquely owned pixels.

In some cases the segments produced at detection may be too large and so over-group sources. While this is not a problem scientifically, it increases the computational complexity of the fit which can lead to poor model performance, or worse, the joint model may even fail to converge altogether. However, unlike morphological dilation which cannot destroy groups of pixels, morphological erosion can destroy the smallest segments typically containing one source. This is a limit that must be avoided in order for \farmer{}'s decision tree to work. More work is required to address this case. 

A limitation of this approach is that groups are defined based on the detection image, its effective resolution, as well as the depths and properties of its constituent bands. A group determined from well-resolved optical images will likely miss pixels with significant flux when applied to sources in drastically lower resolution images. This can be overcome by further dilating each group on a band-by-band basis such that all of the relevant pixels are now constraints on the model. This comes with a dilemma, however, as crowding is worse in low resolution images of the same depth and so light from sources not originally included in the group may now contribute. Yet these nearby sources are not described by the existing group model, and so leaving their flux unaccounted for may instigate a bias in the photometry. The only tractable option seems to be to join these groups and perform the forced photometry in a simultaneous optimization. However, the shapes of these models were n
ever determined together, and so it is uncertain how well the new group of models would perform. Worse, most sources in the deepest IRAC images are blended to some degree and so strictly keeping to this philosophy of joint optimization of all overlapping sources would require every source to be simultaneously fit. While not impossible (e.g. \citealt{Lang2016}), it is potentially computationally expensive. Alternative strategies will be explored in future work.

\textit{What sets the overlap sizes for groups?}
Although groups of sources are limited to their footprint whose pixels are identified by dilating source segments, the groups themselves are saved in memory as rectangular arrays whose dimensions are set by the maximum extent of the group footprint. Although pixels inside of the rectangular array but outside the group footprint (which can often be fractal-like in shape) do not provide any constraining power as their weight is set to zero, the models are still realized onto the larger array during the optimization. It is generally best if these models are not truncated whatsoever, and so \farmer{} enlarges the dimensions of the group array by a set number of pixels. This is not only for numerical reasons internal to \tractor{}, but also is a requirement if post-processing apertures are to measure the full extent of the joint model image. Truncation of that joint model will mean that the wings will not be realized and so the largest apertures will underestimate the true flux. Thankfully, if the models are correctly normalized then the truncation will \textit{not} affect the best-fit normalization coefficient from which the source flux from \tractor{} is derived. Nonetheless it is advisable that the group array size is large enough so that the PSF stamp would not be truncated for a source placed near the edge of main group footprint.

\subsection{Selection Functions and Image Depth}
One must be cognisant regarding which band should be used to determine the models and their best-fit parameters. In fact, this is not a free choice. Using a band outside of those used in the detection image is inappropriate because there may be sources identified in the detection image that do not have flux in the chosen modelling image. If one is to maintain the selection function constructed by the detection strategy, then it must be guaranteed that a detected source has sufficient signal to constrain its model. Otherwise sources without models cannot be photometered, and so the selection function changes in a non-trivial way. For the same reason, it is also inadvisable to model using only one band of a multi-band detection image, or to use just the bands that define the spectral domain of the detection image. Nor is it advisable to attempt to model sources in a co-added image as the effective PSF is not easily characterized, and the FWHM of the constituent PSFs can produce additional variation in the surface brightness profiles. Therefore, it is strongly recommended that the models be produced from precisely the same bands and images that were used or combined to make the detection image. 

Measuring photometry of a source in an image that contains additional sources outside the selection function (by virtue of not being detected) presents another often encountered dilemma, although common also to aperture-based methods. This is because signal from an additional, undetected source is not described by the set of models assigned to a group. For instance, a red source which is undetected in a predominantly blue selection function may in a red band appear next to a known blue source. Although fixing model shapes helps avoid contamination, it is possible that the likelihood will be maximized by increasing the flux parameter of the blue model such that some of the flux from the new, red source is inadvertently accounted for, thus biasing the photometry for the blue source in that red band. Often times these cases can be identified afterwards from diagnostics provided by \farmer{}, although not guaranteed. 

A similar situation is encountered when forcing photometry onto deeper bands of the same wavelength as the detection, and although such images typically can provide better photometric constraints, they may at the same time introduce bias by introducing new, undetected sources. This means that ideally all sources in a forced photometry image should be modelled, which requires that they were detected. However, identifying these new sources automatically ahead of photometry is not practical as lists of detected sources will differ due the blends; the two catalogs must then somehow be reconciled and segmentation maps merged. Doing so in limited numbers is possible with careful supervision, typically with the assumption that new sources are unresolved to avoid re-processing the decision tree (e.g., as used to photometer optically dark sources in \citealt{Jin2022}). This potentially pathological issue will be addressed in future work. 

\subsection{Models, Morphological Corrections, and Drifting}
\label{sec:limit_models}

One significant complication with \farmer{} is that the decision tree needs to be tuned. Because the central operation of the decision tree is to separate resolved and unresolved sources, its parameters are most sensitive to the resolution of the image. Subtended size correlates strongly with apparent brightness, and so sources in deep images typically become unresolved around a certain magnitude threshold. In order to succeed, the decision tree needs to be tuned such that it correctly assigns unresolved models to essentially all sources fainter than this limit, in addition to bright point-sources. A photometric bias can develop if instead the decision tree assigns resolved models to unresolved sources, or vice versa. This can be readily diagnosed from number counts which should be smooth and increase monotonically with decreasing brightness. If the decision tree is not providing adequate model type assignments, the number counts of the detection bands will either contract towards a sharp rise or flatten into a plateau around the resolution threshold. An unresolved model assigned to a resolved source tends to produce an underestimated flux, thereby moving these typically bright but comparatively rare sources towards fainter magnitudes thus creating a plateau. In this case it is likely that the decision tree poorly tuned such that PointSource models are too easily assigned, and so the $\chi^2$ penalty to the PointSource models should be lowered. A resolved model assigned to an unresolved source tends to produce overestimated flux, thereby moving these typically faint but abundant sources towards brighter magnitudes thus creating a sharp rise in counts. In this case the $\chi^2$ penalty of PointSource models should be increased so that it is easier for sources to be assigned an unresolved model. Number counts are not as sensitive to which resolved model is assigned to a resolved source (e.g., ExpGalaxy or DevGalaxy) and so the corresponding parameters are most easily tuned by examining residuals of bright, resolved sources.

It may not be possible to assign a simple parametric model to a particular source. It might be that the source is actually two blended together. Meanwhile, the brightest sources tend to be resolved and have features such as spiral arms, bars, and starbursts that are not described by the smooth models from \tractor{}. As such, model performance tends to decrease for bright, resolved sources (e.g. spirals). {While the presence of poorly fit morphological features will be indicated in the residual statistics (e.g. $\chi^2$), the associated photometry will likely be biased in some way. This is especially true for space-based imaging (e.g., \textit{HST}) where the space spanned by models from \tractor{} are divorced from the real space spanned by highly resolved galaxies. While aperture photometry should be less biased, they are unable to inform about the presence of morphological features.

Chromatic changes in morphology presents a challenge for \farmer{}. The model for a given source during the modelling stage may be simultaneously constrained by multiple bands, but \tractor{} allows only one shape shared between the bands. Therefore the shapes reported by \farmer{} from the modelling stage are most appropriate for the modelling bands with the largest weights. Forced photometry in regular operation proceeds by only allowing the flux to vary with the shape fixed, meaning that changes in morphology are not accounted for by the model. However, \farmer{} makes it possible to perform forced photometry on each band separately so that the shape parameters can be allowed to vary in each case with or without priors, albeit at greater computational expense and runs the danger of over-fitting. It is important to note that forcing models derived from well-resolved bands onto images of lower resolution is typically successful as the larger PSF of the forced photometry band makes the photometric measurement less sensitive to morphology. However, forcing models derived from low resolution images onto bands at high spatial resolution typically results in a poor performance as the band of interest contains more information than the model can describe. 

As discussed in \citet{2020AJ....159..165P}, flux and shape estimates can suffer from biases introduced from inadequate centroiding. Given the great number of multi-wavelength images and facilities involved in modern surveys, even small astrometric offsets can impact the measurements derived from model fitting. Hence, \farmer{} allows the user to unfix the centroid position of each model and introduces a Gaussian prior on its position, on a band-by-band basis. This prior acts to penalize the likelihood of the model fit if the model obtains a centroid that is beyond the distance set by the prior (i.e. `drifts'). This drifting can be especially prevalent in the case of a known faint source next to an undetected bright neighbor which because it is not accounted for by the model will cause the model of the faint source to move towards the bright source, whose unaddressed presence counts against the likelihood more than the original, fainter source. Priors can be set on the position, although their widths are usually determined by successive trials.

It is important to appreciate that the grouping of sources imparts a significant advantage over fitting individual sources. Because groups of sources are photometered separately from other groups, a failure of the model in one group does not affect any other group. Let us consider the unfortunate example in which a galaxy is assigned an inaccurate model whose large axis ratio results in wings extending well beyond the group. While those wings will be a problem for the source in question, and perhaps its group members as well, they will not affect any other group in the image. Hence, while this is an issue for the residual map, there is no reason to be concerned about the photometry of the other group as they were fitted in an entirely separate optimization in isolation of the problematic source. However, this advantage effectively decouples the reconstructed brick-level residual image from the photometry and so complicates searches for sources in residuals. As mentioned in Section~\ref{sec:image_prep}, \farmer{} has built-in functions to filter out these problematic models. 

\subsection{Source De-blending}
\label{sec:deblend}

\begin{figure*}[t]
	\centering
	\includegraphics[width=1\hsize]{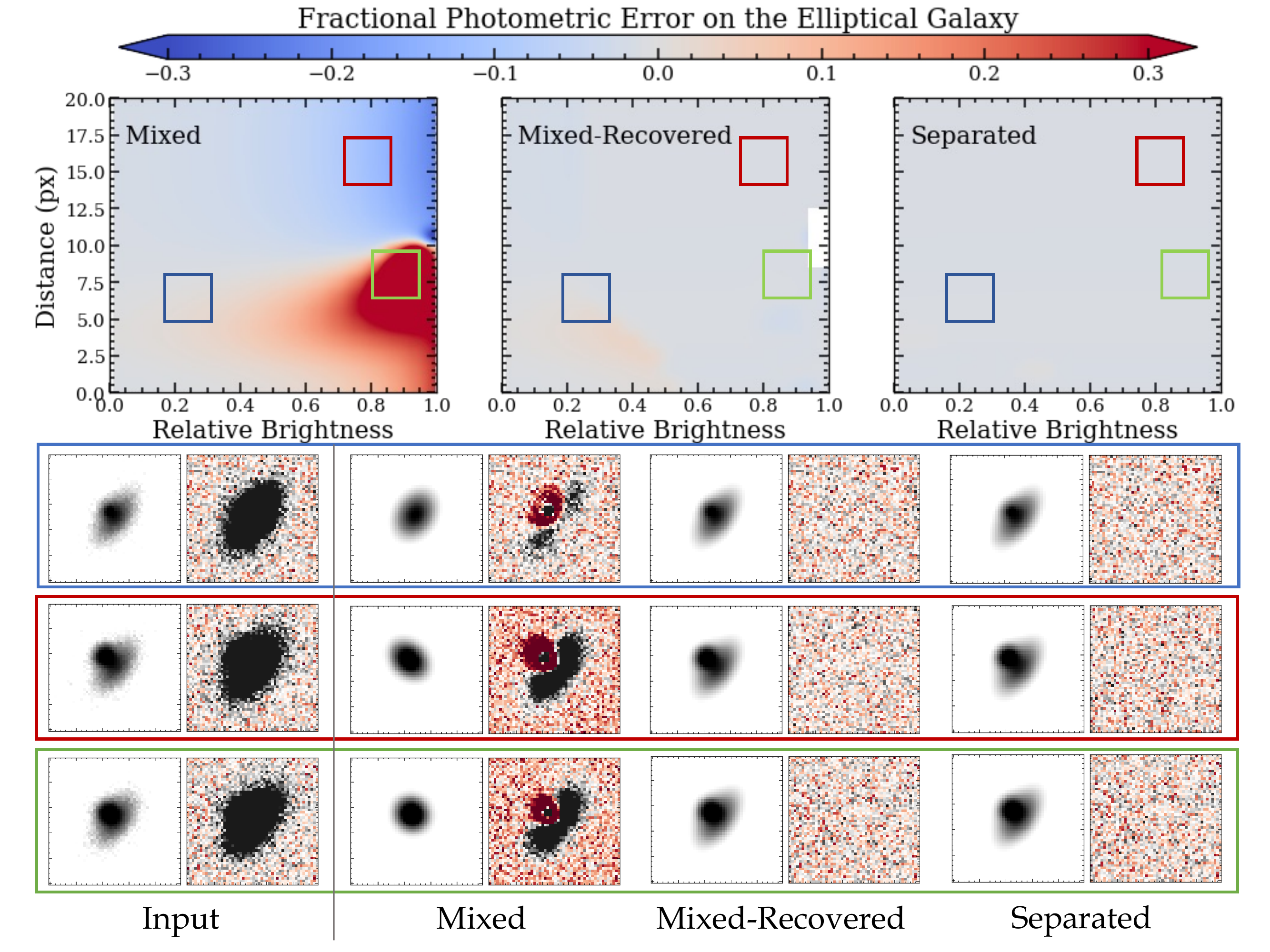}
	\caption{A point source is simulated in the vicinity of a large, central elliptical galaxy. Models are fitted for three cases: the sources are blended and have only one centroid (Mixed), the sources are blended, modelled, and then the missing source is recovered from the residual image and modelled (Mixed-Recovered), and lastly both sources are a priori known and simultaneously fit (separated). Each measurement is repeated over a grid of relative brightness (0 meaning that the point source is negligable) and the distance from the central elliptical to the point source (20 means the point source is in the top left corner). While grey areas indicate successful recovery of the input (true) flux of the elliptical source, red areas indicate that the flux of the elliptical is underestimated. White areas in the middle panel indicate where the point source is not detected in the mixed residual. The model and residual of three situations are shown for each measurement strategy. 
	}
	\label{fig:demo_pointexp}
\end{figure*}

While profile-fitting photometry can be used to de-blend two sources, they first must be identified as separate objects, which in turn depends on the original source detection. As such, de-blending sources at the detection stage is not a problem which profile-fitting photometry can (or should) solve and instead is well-suited to address the related, but distinct issue of accurately measuring the flux of two \textit{identified} but blended sources. It is essential, therefore, to understand that if two nearby sources are not successfully de-blended at detection, then profile-fitting techniques should not be expected to reliably de-blend them either.

This concept is demonstrated in Figure~\ref{fig:demo_pointexp} whereby a point source is placed in the vicinity of a bright resolved galaxy and appropriate models are assumed to be known. Several cases are set up by varying their distance and relative brightness. Attempting to photometer both of them with only one model produces expectedly poor fits in several cases. The system is then evolved by allowing it to subtract the first source, find the brightest residual source, and re-fit the original using two centroids which in turn improves the performance in cases where the residual source can be identified. However, in practice this is dangerous if one does not know beforehand whether there is another source or whether the model for the one source was simply a poor fit. Lastly, the two sources are fit by two appropriate models which results in accurate photometry at all distances and relative brightness. Hence, de-blending with profile-fitting photometry requires the correct number of models (and centroids) to optimize for a given group of sources.

What is little appreciated, however, is that this behavior is undoubtedly an advantage. While aperture techniques do not make any assumption about source morphology and are hence extremely powerful in the face of resolved structural features in galaxies, their ability to identify cases of sources blended at detection or quantify contamination in photometry of photometrically blended sources is severely limited. Parametric profile-fitting techniques suffer from neither of these drawbacks. So long as intrinsically blended sources are not well-described also by a single profile\footnote{In such cases sources cannot be identified as separate objects without higher resolution ancillary data anyways.}, then the optimized model will not achieve a satisfactory fit. These cases may be confidently identified a posteriori using statistics such as those discussed in Section~\ref{sec:catalogs}.

\subsection{Comparison to Similar Methods}
\label{sec:comparison}

These advantages and limitations hold mainly for purely parametric models. There exists another, related class which uses a high resolution cutout of a given source as its model that can be used to photometer other bands by first convolving it with an appropriate kernel to translate its native PSF to that of the band of interest, and then scaling the unit normalized model to match the source in that band. These `stamps' have a distinct advantage over parametric models in that they can exploit the resolution of the cutout image to capture structural features not describable by typically smooth parametric models. Without shape parameters to constrain, these stamps can be extremely efficient in measuring fluxes as essentially a scaling factor between the PSF-transformed stamp and the source in question. While simpler than purely parametric models, this approach requires a deep high-resolution image which contains every detected source (if not the same image as that used for detection). More so, the PSF must be well-understood to provide kernels to map the original PSF to those of the lower resolution images, a drawback not shared by parametric models. The stamp must also be resampled to match the pixel scale of the image to be photometered. For example, an \textit{HST}-derived stamp of a marginally resolved source applied to \textit{Spitzer} provides no significant advantage over a parameteric model. Worse, the morphology described by the stamp is assumed to be constant, and so there can arise significant effects between the wavelength of the stamp image and that of the image to be photometered. Such stamp-based profile-fitting software include \texttt{TFIT} \citep{laidler07_tfit}, \texttt{T-PHOT} \citep{2015A&A...582A..15M,merlin16_tphot},  \texttt{PyGFIT} \citep{2013PASP..125.1514M}, \texttt{Morfometryka} \citep{Ferrari2015}, \texttt{LAMBDAR} \citep{2016MNRAS.460..765W}, and GOLFIR \citep{Kokorev2022}. Each one takes its own approach to the problem of flu
 x estimation in terms of available models, parametrization, algorithm speed, flexibility, and accessibility. As discussed in Section~\ref{sec:limit_models}, purely parametric models can overcome the limitations of these stamp-based codes by freely fitting the shape of the model, possibly with some prior constraints (e.g., \texttt{GALFIT}, \citealt{Peng2002, Peng2010}, \texttt{ProFit}, \citealt{Robotham2017}; \texttt{SExtractor++}, \citealt{Bertin2020, Kummel2020}; \texttt{GALAPAGOS-2}, \citealt{Haussler2022}).

One of the most similar photometry frameworks to \farmer{} is \texttt{HSCPipe}, in part because they both inherit the profile-fitting approach of SDSS \citep{stoughton_2002}. As discussed in \citet{aihara_second_2019}, \texttt{HSCPipe} provides model-based photometry by fitting both point-source (PSF) and composite galaxy (cModel) profiles to each galaxy individually. Even though both resolved and point-like models are tried, unlike \farmer{} they are tried for each source independent of their neighbors, which for blended sources can lead to inconsistencies (as demonstrated in Figure~\ref{fig:demo_photmethods}). Furthermore, \texttt{HSCPipe} does not choose a best-fit model type for each source and instead provides fluxes measured from each profile assuming independence from neighbors. While this is computationally faster than a decision tree, it is also inefficient to fit unresolved sources with highly parameterized composite models (which risk overfitting). As of version 8 of \texttt{HSCPipe}\footnote{\url{https://hsc.mtk.nao.ac.jp/pipedoc/pipedoc_8_e/index.html}}, only likelihood of the CModel fits are reported and so a consistent statistical comparison with the PSF models is not possible, which leaves only a binary extendedness flag to indicate a resolved source. \farmer{} provides not only a best-fit model type for each source, but also suite of statistics from which the reliability of that model can be assessed. 

Although limited to low resolution IRAC images, the \texttt{IRACLEAN} software \citep{Hsieh_2012} measures photometry by iteratively subtracting PSFs at detected source centroids until the residual is clean of signal to some user defined level. Although broadly similar to \farmer{}, \texttt{IRACLEAN} does not perform model-fitting in a classical sense as an unbounded iterative subtraction of the PSF is equivalent to an model with effectively unlimited parameters. Furthermore, the order in which sources are processed can introduce hysteresis in crowded regions. There is also the danger of overfitting, as \texttt{IRACLEAN} will continue subtracting a scaled PSF stamp until a given segment has no more signal, which in the case of a blend will combine the flux of the two sources into one photometric measurement. \farmer{}'s parametric models act as a prior which can, in some cases, ignore the flux of a neighbor which is left in the residual, and report statistics flagging the problem to the user. Further discussions and comparisons with \texttt{IRACLEAN} are presented in \citet{Weaver2022_catalog}.

\subsection{Computational Considerations}
\label{sec:computation}

Computation of sources scales with the number of sources fit simultaneously as well as the number of free parameters, meaning that these techniques require significantly longer runtimes compared to aperture photometry. 
In the context of modern deep surveys containing millions of sources (many of which are resolved), fitting all sources simultaneously would be enormously complex requiring significant computational resources. However, a high degree of parallelization can be achieved so long as the source density and resolution allow for distinct groups of sources to be identified and fit separately. A practical approach is to process each brick independently. Source groups are constructed at runtime and kept in memory only, so they are ideal for being run in parallel, e.g. across many CPUs of a given cluster node. 

However, computational time still increases with the number of free parameters. As such, the modelling stage is not only more complicated than forced photometry because of the several trials of the decision tree, but also because shapes are left to vary in some stages. It is for this reason that the decision tree starts with simple models and moves towards complexity, or in other words, computational expense. If the conditions of the decision tree are satisfied for every source, then the models are assigned without moving to the next stage. For example, an isolated point source should only be tried out with a PointSource and SimpleGalaxy model whereupon it should satisfy the PointSource criterion and stop. Each of these model types have three parameters (two for position and one for flux) and so are incredibly quick compared to a CompositeGalaxy with ten parameters.

Unfortunately, computational time also increases with source crowding given the stronger covariance between neighboring models. Many separate sources can be modelled independently and in parallel without a loss of accuracy. However, because deep images of crowded fields are best photometered when groups of nearby sources are simultaneously modelled, the complexity and computational expense is greater than if the same number of sources were fit separately. As such, it is strongly advised that typical source groups contain as few members as possible without breaking across two blended sources. Unfortunately, the situation of source crowding will only become more difficult as surveys grow deeper. While apertures will eventually hit a limit, profile-fitting photometry can forge ahead, albeit with a greater computational cost.

\end{document}